\documentclass[a4paper,11pt]{article}
\usepackage{jheppub}

\usepackage[utf8]{inputenc}
\usepackage{graphicx,cancel,float}
\usepackage{subfigure}
\usepackage{amssymb}
\usepackage{textcomp}
\usepackage{amsmath}
\usepackage{tabularx}
\usepackage{bm}
\usepackage{times}
\usepackage{color}
\usepackage{slashed}
\usepackage{multirow}
\usepackage{verbatim}
\usepackage{epsfig,epstopdf}

\allowdisplaybreaks[4]

\def\lsim{\mathrel{\raise.3ex\hbox{$<$\kern-.75em\lower1ex\hbox{$\sim$}}}}
\def\gsim{\mathrel{\raise.3ex\hbox{$>$\kern-.75em\lower1ex\hbox{$\sim$}}}}

\definecolor{orange}{rgb}{1,0.5,0}

\preprint{}

\title{Constraints on general neutrino interactions with exotic fermion from neutrino-electron scattering experiments}
\author[a]{Zikang Chen,}
\author[b]{Tong Li,}
\author[a]{Jiajun Liao}
\affiliation[a]{School of Physics, Sun Yat-Sen University, Guangzhou 510275, China }
\affiliation[b]{School of Physics, Nankai University, Tianjin 300071, China}
\emailAdd{chenzk7@mail2.sysu.edu.cn}
\emailAdd{litong@nankai.edu.cn}
\emailAdd{liaojiajun@mail.sysu.edu.cn}

\abstract{
The couplings between the neutrinos and exotic fermion can be probed in both neutrino scattering experiments and dark matter direct detection experiments. We present a detailed analysis of the general neutrino interactions with an exotic fermion and electrons at neutrino-electron scattering experiments. We obtain the constraints on the coupling coefficients of the scalar, pseudoscalar, vector, axialvector, tensor and electromagnetic dipole interactions from the CHARM-II, TEXONO and Borexino experiments. For the flavor-universal interactions, we find that the Borexino experiment sets the strongest bounds in the low mass region for the electromagnetic dipole interactions, and the CHARM-II experiment dominates the bounds for other scenarios. If the interactions are flavor dependent, the bounds from the CHARM-II or TEXONO experiment can be avoided, and there are correlations between the flavored coupling coefficients for the Borexino experiment. We also discuss the detection of sub-MeV DM absorbed by bound electron targets and illustrate that the vector coefficients preferred by XENON1T data are allowed by the neutrino-electron scattering experiments.
}

\arxivnumber{arXiv:2102.09784 }

\begin{document}

\maketitle
\setcounter{page}{2}

\newpage

\section{Introduction}
\label{sec:Intro}
The phenomenon of neutrino oscillations has been well confirmed by various neutrino experiments in the last two decades~\cite{Zyla:2020zbs}. Since the explanation of neutrino oscillations requires nonvanishing neutrino masses, which cannot be accounted for by the Standard Model (SM), the observation of neutrino oscillations provides a strong motivation to search for new physics beyond the SM that are associated with neutrinos.
Moreover, the existence of dark matter (DM) through abundant cosmological and astrophysical
observations is one of the most plausible evidences of new physics beyond the SM.
The DM direct detection experiments have pushed the limit on the cross section of DM scattering off nucleus close to
the neutrino floor for weak scale DM. The couplings between the neutrinos and sub-GeV DM through the scattering off nucleus have been studied in Ref.~\cite{Brdar:2018qqj, Chang:2019sel, Dror:2019onn, Dror:2019dib, Chang:2020jwl, Li:2020lba, Hurtado:2020vlj, Li:2020pfy}. There are also plenty of novel models and signatures proposed to search for sub-GeV DM through the scattering off electrons~\cite{Ge:2020jfn,Shoemaker:2020kji,Shakeri:2020wvk, Dror:2020czw, Brdar:2020quo,AristizabalSierra:2020zod}.

Recently, the XENON collaboration reported an excess of electronic recoil events with the energy around 2-3 keV~\cite{Aprile:2020tmw} and the event distribution has a broad spectrum for the excess. They collected low energy electron recoil data from the XENON1T experiment with an exposure of 0.65 tonne-years and analyzed various backgrounds for the excess events. Although a small tritium background fits the excess data well, the solar axion explanation or the solar neutrinos with magnetic moment can also provide a plausible source for the peak-like excess. However, both of the two scenarios have tension with stellar cooling constraints~\cite{Viaux:2013lha,Diaz:2019kim,DiLuzio:2020jjp,Gao:2020wer,Dent:2020jhf,Brdar:2020quo}. Some studies instead proposed to explain the XENON1T excess through the electron recoil by solar neutrinos with the sterile neutrino DM in the final states of inelastic scattering~\cite{Ge:2020jfn,Shoemaker:2020kji}. On the other hand, the inverse process in which the incoming fermionic DM is absorbed by bound electron
targets and emits a neutrino is sensitive to the DM with mass below MeV~\cite{Shakeri:2020wvk,Dror:2020czw}. The two kinds of signals are governed by the same interactions between SM neutrino and the exotic fermion. The relevant interactions are inevitably constrained by the precision measurements in neutrino experiments~\cite{Brdar:2020quo,AristizabalSierra:2020zod}. In this work we study the constraints on general neutrino interactions with sub-GeV exotic fermion from neutrino-electron scattering experiments.

The large volume detectors enable precise measurements of neutrino properties. The large neutrino detectors like Borexino can be used to place constraints on general neutrino interactions. The Borexino experiment, located at the Laboratori Nazionali del Gran Sasso, was built with a primary goal of measuring solar neutrinos from the $pp$ chain~\cite{Agostini:2017ixy}. We employ the Borexino measurements of low energy solar neutrinos to set limits on the neutrino-electron scattering with an outgoing fermion $\chi$
\begin{eqnarray}
\nu e\to \chi e\;,
\end{eqnarray}
where $\chi$ could be sterile neutrino or other possible exotic fermions. The results apply for the exotic fermion $\chi$ being either DM candidate or not. We restrict the general neutrino interactions categorized by dimension-5 dipole operators and dimension-6 four-fermion operators. They respect Lorentz invariance and the gauge symmetries $SU(3)_c\times U(1)_{\rm em}$. The scattering cross section from the magnetic and electric dipole operators is inversely proportional to the recoil energy and thus the experiments with low energy threshold are sensitive to them. For the four-fermion interactions, all Lorentz-invariant operators (scalar, vector, pseudoscalar, axialvector and tensor) will be explored in the neutrino-electron scattering. As the produced solar electron neutrinos oscillate into muon and tau neutrinos, we can also place limits on the general interactions of all neutrino flavors. In addition, accelerator neutrinos with the energy being several tens of GeV can be used to exploit large $\chi$ mass region. We thus take into account the constraints from the CHARM-II experiment~\cite{Vilain:1993kd, Vilain:1994qy} as well as reactor neutrino using TEXONO~\cite{Deniz:2009mu} data. The relevant data were used to study the electromagnetic interactions or the nonstandard vector-type interactions of a neutrino via the transition $\nu e\to \nu e$~\cite{Khan:2014zwa,Khan:2016uon,Khan:2017oxw,Khan:2017djo,Lindner:2018kjo,Khan:2019jvr}.
	The heavy neutral leptons as well as DM scattering off electrons can be also searched in high-energy neutrino beam-dump experiment or far-forward detector at the LHC~\cite{Jodlowski:2020vhr,Batell:2021blf}.

This paper is organized as follows. In Sec.~\ref{sec:Model}, we discuss the effective Lagrangian of an exotic fermion interacting with neutrino and electron. Then we display the amplitudes and differential cross sections of neutrino-electron scattering with the outgoing exotic fermion. In Sec.~\ref{sec:experiments}, we consider three neutrino-electron scattering experiments: CHARM-II, TEXONO and Borexino, and show the constraints on the general neutrino interactions.
The detection of sub-MeV DM absorbed by bound electron targets is then discussed in Sec.~\ref{sec:DM}. Finally, we summarize our conclusions in Sec.~\ref{sec:Con}.

\section{The general neutrino interactions with exotic fermion via neutrino-electron scattering}
\label{sec:Model}

\begin{figure}[htb]
	\centering
	\includegraphics[width=0.45\textwidth]{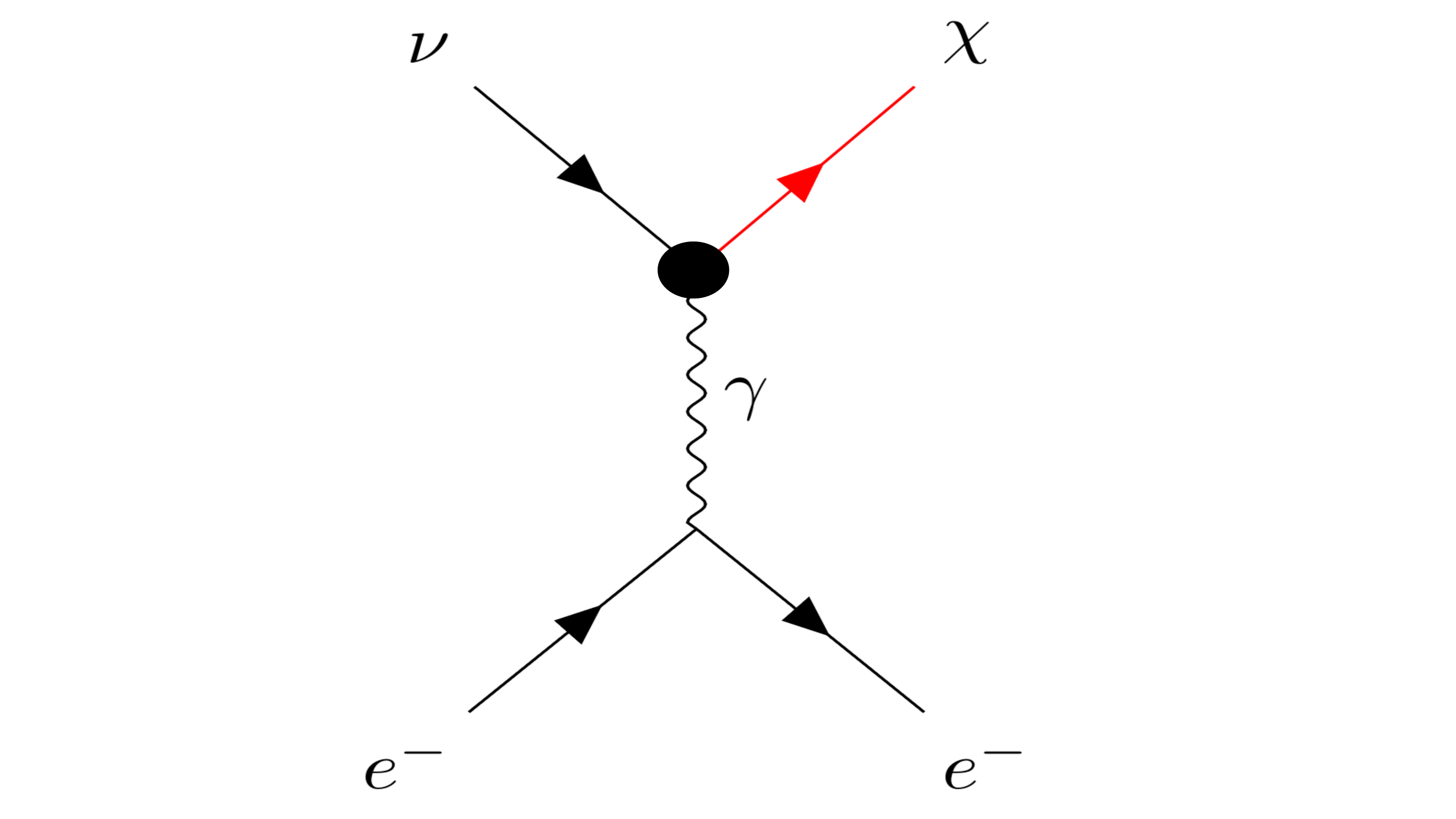}
	\includegraphics[width=0.45\textwidth]{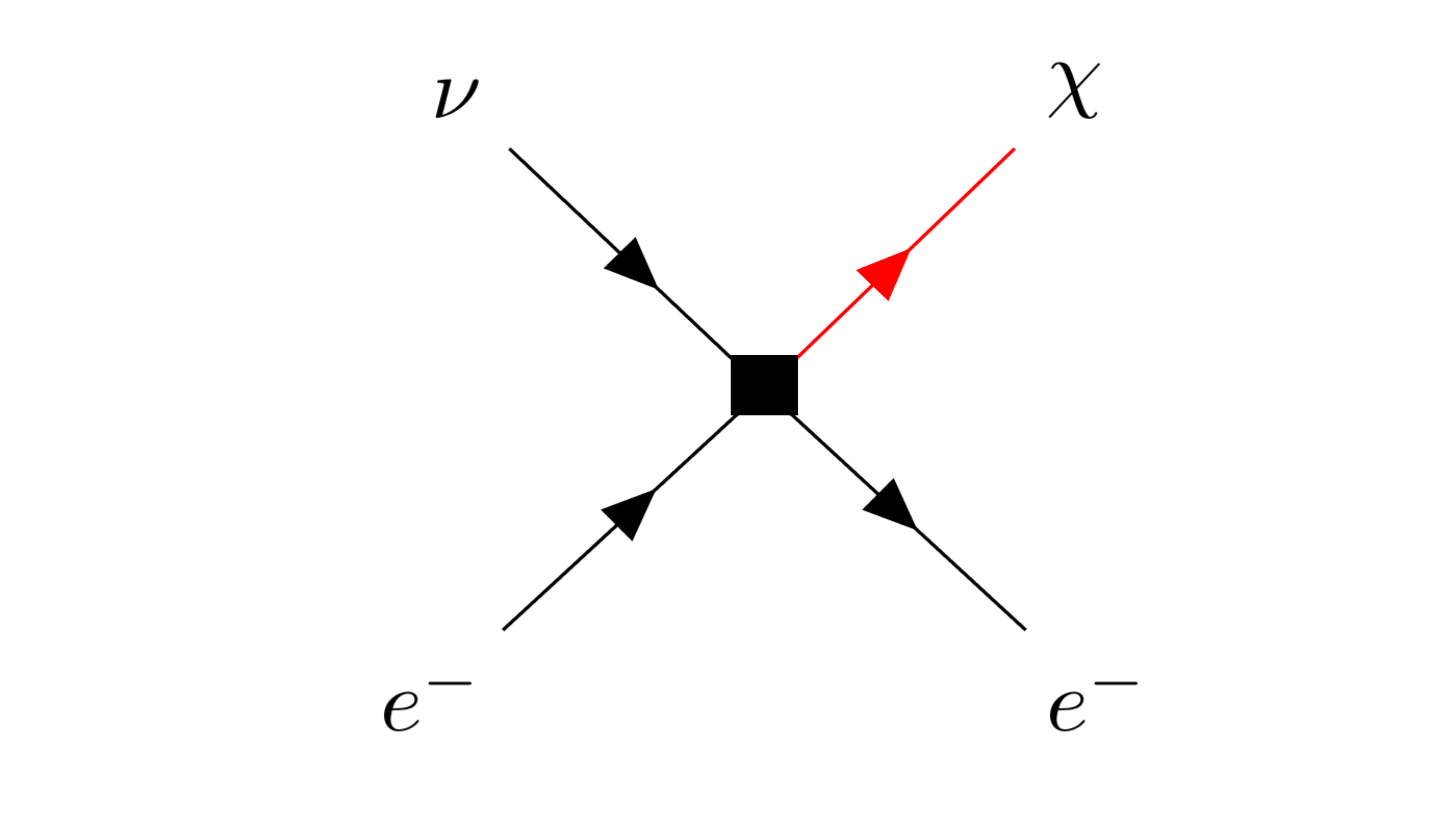}
	\caption{The tree-level Feynman diagrams for the $\nu_\alpha +e^- \rightarrow \chi +e^-$ process, where the circular bulb (square) represents the effective dim-5 dipole (dim-6 four-fermion) interaction.}
	\label{fig:feyndiag}
\end{figure}

We consider a Dirac fermion $\chi$ and its general interactions with neutrino and electron.
The effective Lagrangian including both dim-5 dipole operators and dim-6 four-fermion operators reads as
\begin{eqnarray}
\mathcal{L}\supset {G_F\over \sqrt{2}}\left[\sum_a\bar{\chi} \Gamma^a \nu_\alpha~\bar{e} \Gamma^a(\epsilon^a_\alpha + \tilde{\epsilon}^a_\alpha \gamma^5) e +{v_H\over \sqrt{2}}\bar{\chi}\sigma^{\mu\nu}(\epsilon^M_\alpha+\epsilon^E_\alpha \gamma_5)\nu_\alpha F_{\mu\nu}\right] + h.c.\;,
\label{eq:Leff}
\end{eqnarray}
where $\alpha\equiv\{e, \mu, \tau\}$, $v_H\simeq 246$ GeV is the vacuum expectation value of the SM Higgs, $F_{\mu\nu}$ is the electromagnetic field strength tensor and $\Gamma^a\equiv \{I, i\gamma^5, \gamma^\mu, \gamma^\mu \gamma^5, \sigma^{\mu\nu}\equiv{i\over 2}[\gamma^\mu,\gamma^\nu]\}$ correspond to the scalar ($S$), pseudoscalar ($P$), vector ($V$), axialvector ($A$) and tensor ($T$) operator, respectively. The four-fermion operators are analogous to those in Ref.~\cite{Khan:2019jvr}. Here the dimensionless parameters $\epsilon_\alpha^M$, $\epsilon_\alpha^E$, $\epsilon_\alpha^a$ and $\tilde{\epsilon}_\alpha^a$ are in general complex.
The presence of new interactions of Eq.~(\ref{eq:Leff}) will give rise to the tree-level neutrino-electron scattering, as shown in Fig.\ref{fig:feyndiag}.

In the SM, the neutrino-electron scattering is governed by both the weak neutral current (NC) and charged current (CC). The effective Lagrangian for the SM NC is given by
\begin{eqnarray}
\mathcal{L}_{\rm NC}={G_F\over \sqrt{2}}\bar{\nu}\gamma^\mu (1-\gamma_5)\nu \bar{e}\gamma_\mu (g_V-g_A\gamma_5) e\;,
\end{eqnarray}
where $g_V=-{1\over 2}+2\sin^2\theta_W$ and $g_A=-{1\over 2}$. The CC Lagrangian can be transmitted as
\begin{eqnarray}
\mathcal{L}_{\rm CC}={G_F\over \sqrt{2}}\bar{\nu}\gamma^\mu (1-\gamma_5)\nu \bar{e}\gamma_\mu (1-\gamma_5) e\;.
\end{eqnarray}
The CC current only contributes to the scattering of $\nu_e$.
The differential cross section of neutrino-electron scattering in the SM is~\cite{Khan:2019jvr}
\begin{eqnarray}
\label{eq:xsectionSM}
\frac{d\sigma^{\rm SM}_{\alpha\beta}}{dE_R} &=& {G_F^2 m_e\over 2\pi} \Big[\Big(g^L_{\alpha\beta}\Big)^2+\Big(g^R_{\alpha\beta}\Big)^2 \Big(1-{E_R\over E_\nu}\Big)^2-g^L_{\alpha\beta} g^R_{\alpha\beta} {m_e E_R\over E_\nu^2}\Big]\;,
\end{eqnarray}
where $\alpha$ ($\beta$) denotes the flavor of the neutrino in the initial (final) states, $E_\nu$ is the neutrino energy, $E_R$ is the electron recoil energy and
\begin{eqnarray}
\Big(g^L_{\alpha\beta},g^R_{\alpha\beta}\Big)=\left\{
\begin{array}{ll}
(2\sin^2\theta_W+1,2\sin^2\theta_W), & \hbox{$\alpha=\beta=e$;} \\
(2\sin^2\theta_W-1,2\sin^2\theta_W), & \hbox{$\alpha=\beta=\mu,\tau$;} \\
0, & \hbox{$\alpha\neq\beta$.}
\end{array}
\right.
\end{eqnarray}

We then calculate the differential cross section of $\nu_\alpha+e \rightarrow \chi + e$ by following the procedure given in the appendix. Here we show the differential cross sections for different operators:
\begin{eqnarray}
\label{eq:xsectionS}
\frac{d\sigma^S_{\nu_\alpha}}{dE_R} &=& \frac{G_F^2m_e}{8\pi}\left[|\epsilon_\alpha^S|^2\left(1+\frac{E_R}{2m_e}\right)+|\tilde{\epsilon}_\alpha^S|^2\frac{E_R}{2m_e}\right]\left(\frac{m_eE_R}{E_\nu^2}+\frac{m_\chi^2}{2E_\nu^2}\right)\;,
\\
\label{eq:xsectionP}
\frac{d\sigma^P_{\nu_\alpha}}{dE_R}&=& \frac{G_F^2m_e}{8\pi}\left[|\epsilon_\alpha^P|^2\frac{E_R}{2m_e}+|\tilde{\epsilon}_\alpha^P|^2\left(1+\frac{E_R}{2m_e}\right)\right]\left(\frac{m_eE_R}{E_\nu^2}+\frac{m_\chi^2}{2E_\nu^2}\right)\;,
\\
\label{eq:xsectionV}
\frac{d\sigma_{\nu_\alpha}^V}{dE_R} &=& \frac{G_F^2m_e}{4\pi}\left[\left(|\epsilon_\alpha^V|^2+|\tilde{\epsilon}_\alpha^V|^2\right)\left(1-{E_R\over E_\nu}+{E_R^2\over 2E_\nu^2}-{m_\chi^2\over 2E_\nu m_e}+{m_\chi^2E_R\over 4E_\nu^2 m_e}\right) \right.
\nonumber\\
& &\left.-\left(|\epsilon_\alpha^V|^2-|\tilde{\epsilon}_\alpha^V|^2\right)\left({E_Rm_e\over 2E_\nu^2}+{m_\chi^2\over 4 E_\nu^2}\right)-2\text{Re}[\epsilon_\alpha^V(\tilde{\epsilon}_\alpha^V)^*]{E_R\over E_\nu}\left(1-{E_R\over 2E_\nu}-{m_\chi^2\over 4E_\nu m_e}\right)\right] \;,
\\
\label{eq:xsectionA}
\frac{d\sigma_{\nu_\alpha}^A}{dE_R} &=& \frac{G_F^2m_e}{4\pi}\left[\left(|\epsilon_\alpha^A|^2+|\tilde{\epsilon}_\alpha^A|^2\right)\left(1-{E_R\over E_\nu}+{E_R^2\over 2E_\nu^2}-{m_\chi^2\over 2E_\nu m_e}+{m_\chi^2E_R\over 4E_\nu^2 m_e}\right) \right.
\nonumber\\
& &\left.+\left(|\epsilon_\alpha^A|^2-|\tilde{\epsilon}_\alpha^A|^2\right)\left({E_Rm_e\over 2E_\nu^2}+{m_\chi^2\over 4 E_\nu^2}\right)-2\text{Re}[\epsilon_\alpha^A(\tilde{\epsilon}_\alpha^A)^*]{E_R\over E_\nu}\left(1-{E_R\over 2E_\nu}-{m_\chi^2\over 4E_\nu m_e}\right)\right] \;,
\\
\label{eq:xsectionT}
\frac{d\sigma_{\nu_\alpha}^T}{dE_R} &=& \frac{2G_F^2m_e|\epsilon_\alpha^{T}-\tilde{\epsilon}_\alpha^T|^2}{\pi}\left[1-{E_R\over E_\nu}+{E_R^2\over 4E_\nu^2}-{E_R m_e\over 4E_\nu^2}-{m_\chi^2\over 4E_\nu^2}\left({1\over2} +{2E_\nu\over m_e}-{E_R\over 2 m_e}\right)\right]\;,
\\
\label{eq:xsectionEM}
\frac{d\sigma_{\nu_\alpha}^\text{EM}}{dE_R} &=& { 2\sqrt{2}\alpha_\text{EM}G_F |\epsilon_\alpha^{E}-\epsilon_\alpha^{M}|^2 \over m_e}
\left[{m_e \over E_R}-{m_e \over E_\nu}-{m_\chi^2 \over E_\nu E_R}\left({1\over 2} -{E_R\over 4E_\nu}+ {m_e\over E_\nu}\right)
-{m_\chi^4\over 8 E_\nu^2 E_R^2}\left(1-{E_R\over m_e}\right)\right] \;,
\end{eqnarray}
where $\alpha_\text{EM}\simeq1/137$ is the electromagnetic fine structure constant. Here we assume each of the scalar, pseudoscalar, vector, axialvector, tensor and electromagnetic dipole operators dominates at a time.
The differential cross sections of $\bar{\nu}_\alpha+e \rightarrow \bar{\chi} + e$ are the same as those of $\nu_\alpha+e \rightarrow \chi + e$ except for the cross term of the axialvector operator in Eq.~(\ref{eq:xsectionA}) changed sign.
From the above equations,
we realize that the the bounds on $\{\tilde{\epsilon}_\alpha^S, \tilde{\epsilon}_\alpha^P, \tilde{\epsilon}_\alpha^V, \tilde{\epsilon}_\alpha^A, \tilde{\epsilon}_\alpha^T, \epsilon_\alpha^{E}\}$ will be the same as $\{\epsilon_\alpha^P, \epsilon_\alpha^S, \epsilon_\alpha^A, \epsilon_\alpha^V, \epsilon_\alpha^{T}, \epsilon_\alpha^{M}\}$ if we consider only one $\epsilon$ parameter at a time.

\section{Neutrino-electron scattering experiments}
\label{sec:experiments}

The parameter space of the general neutrino interactions with electrons can be constrained by various neutrino-electron scattering experiments such as CHARM-II~\cite{Vilain:1993kd, Vilain:1994qy}, LAMPF~\cite{Allen:1992qe}, LSND~\cite{Auerbach:2001wg},  TEXONO~\cite{Deniz:2009mu} and MINER$\nu$A~\cite{Park:2015eqa, Valencia:2019mkf}.
The precision of these neutrino-electron scattering experiments can be inferred from their measurement of $\sin^2\theta_W$~\cite{Rodejohann:2017vup}, i.e., $\sin^2\theta_W=0.2324\pm0.0083$ at CHARM-II, $0.249\pm0.063$ at LAMPF~\cite{Allen:1992qe}, $0.248\pm0.051$ at LSND~\cite{Auerbach:2001wg}, and $0.251\pm0.039$ at TEXONO~\cite{Deniz:2009mu}. Hence, we consider CHARM-II and TEXONO in our analysis since they have the strongest sensitivity to $\sin^2\theta_W$\footnote{The neutrino-electron scattering data at MINER$\nu$A can be also used to impose competitive bounds on new interactions as CHARM-II, see e.g.~\cite{Arguelles:2018mtc}. However, the full analysis of the measurement at MINER$\nu$A requires a dedicated Monte Carlo simulation of the sideband events, which is not available to us.}. We also consider the Borexino experiment since it measures the solar neutrinos which have much lower energies than other experiments.

\subsection{CHARM-II}
\begin{figure}[htb!]
	\begin{center}
		\includegraphics[scale=1,width=0.48\linewidth]{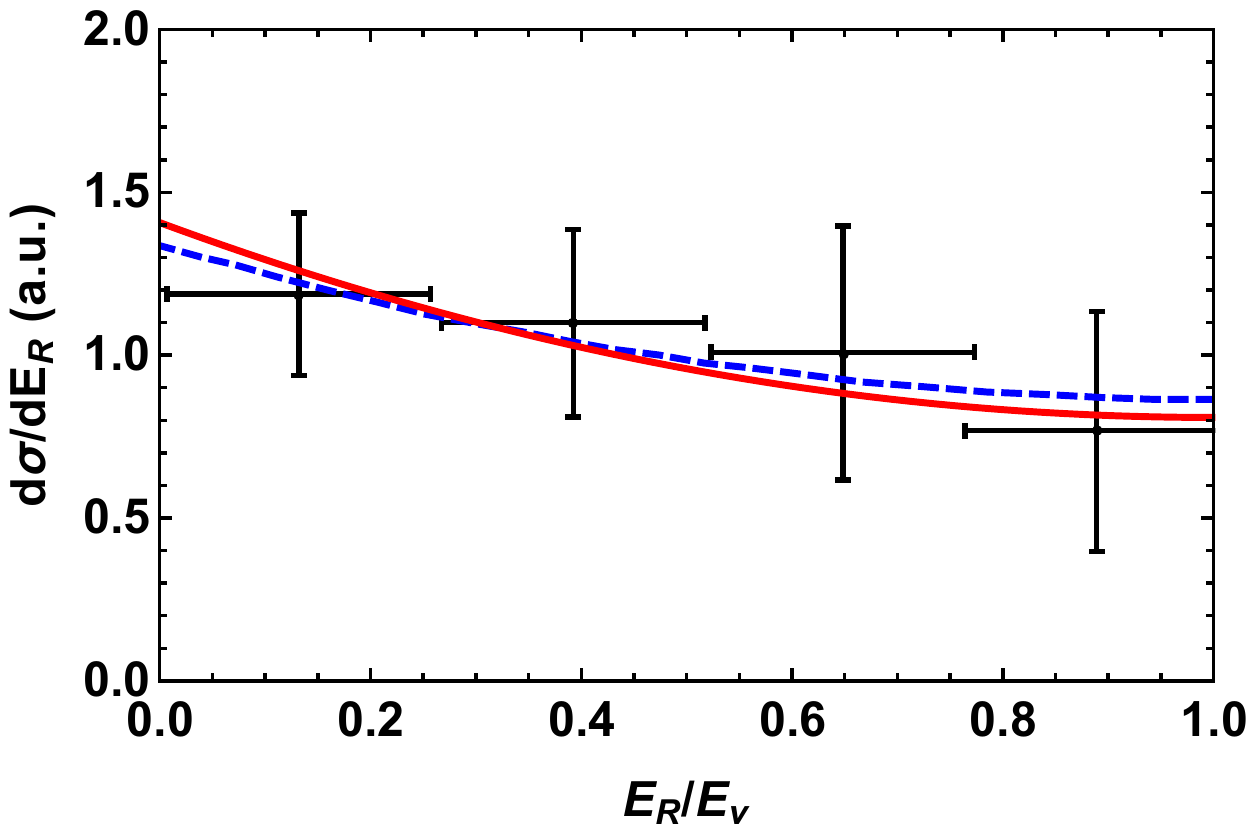}
		\includegraphics[scale=1,width=0.48\linewidth]{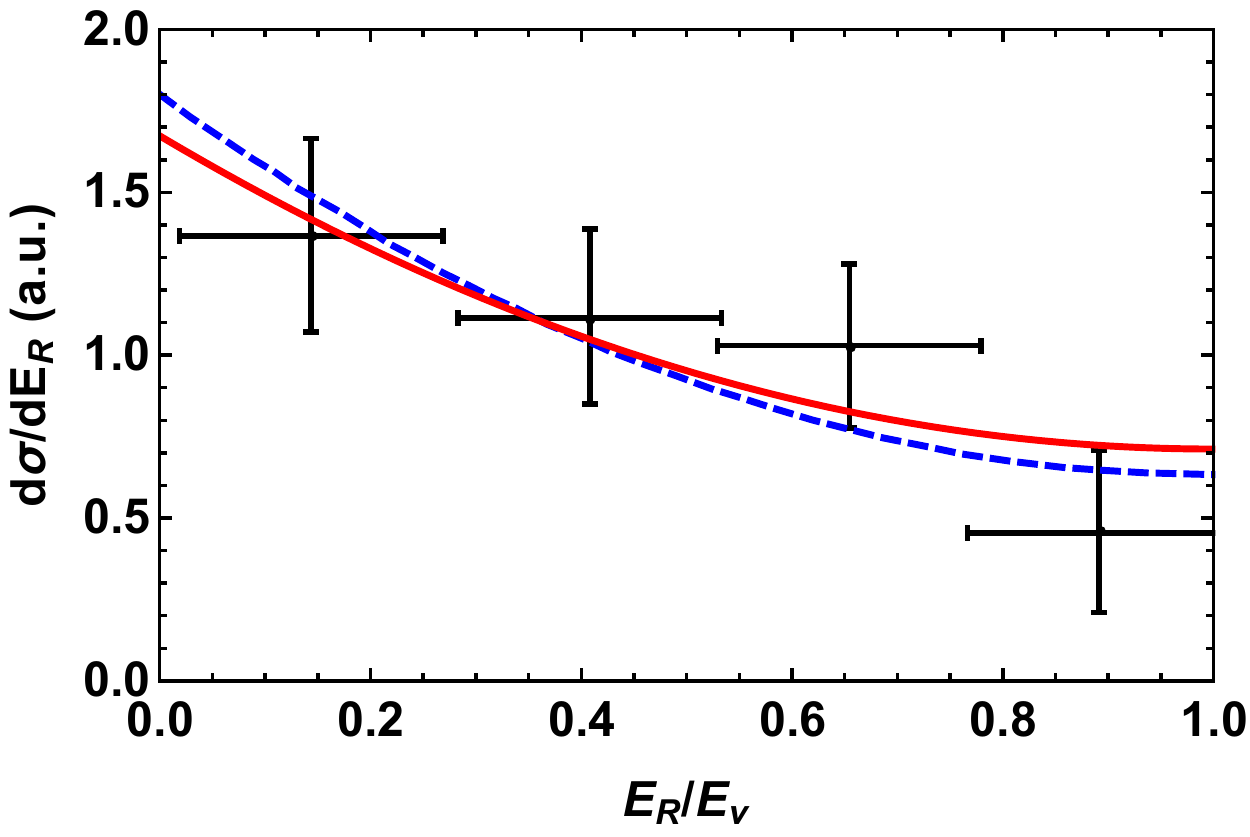}
	\end{center}
	\caption{The differential cross sections of the $\nu_\mu$ (left) and $\bar{\nu}_\mu$ (right) scattering on electrons at CHARM-II. The black data points and the blue dashed SM prediction curve are taken from Ref.~\cite{Vilain:1993kd}. The red curve corresponds to our best-fit prediction in the SM.
	}
	\label{fig:charm}
\end{figure}

The CHARM-II experiments measured the high energy $\nu_\mu$ and $\bar{\nu}_\mu$ beam from the Super Proton Synchrotron (SPS) at CERN~\cite{Vilain:1993kd, Vilain:1994qy}. The mean neutrino energies of the $\nu_\mu$ and $\bar{\nu}_\mu$ beam are 23.7~GeV and 19.1~GeV, respectively. The unfolded differential cross sections from the measurement have been given in Ref.~\cite{Vilain:1993kd}, and the data points are shown in Fig.~\ref{fig:charm}. Our SM predictions are consistent with those given in the Ref.~\cite{Vilain:1993kd}. Therefore, we consider the following $\chi^2$ function in our analysis for new physics:
\begin{eqnarray}
\chi^2_\text{CHARM-II}=\sum_i\frac{(d\sigma/dE_R)_i-s_i^0)^2}{\sigma_i^2}+ (\nu_\mu\rightarrow\bar{\nu}_\mu)\,,
\end{eqnarray}
where $s_i^0$ and $\sigma_i$ are the measured differential cross section and its corresponding uncertainties taken from Ref.~\cite{Vilain:1993kd}.

\subsection{TEXONO}
\begin{figure}[htb!]
	\begin{center}
		\includegraphics[scale=1,width=0.68\linewidth]{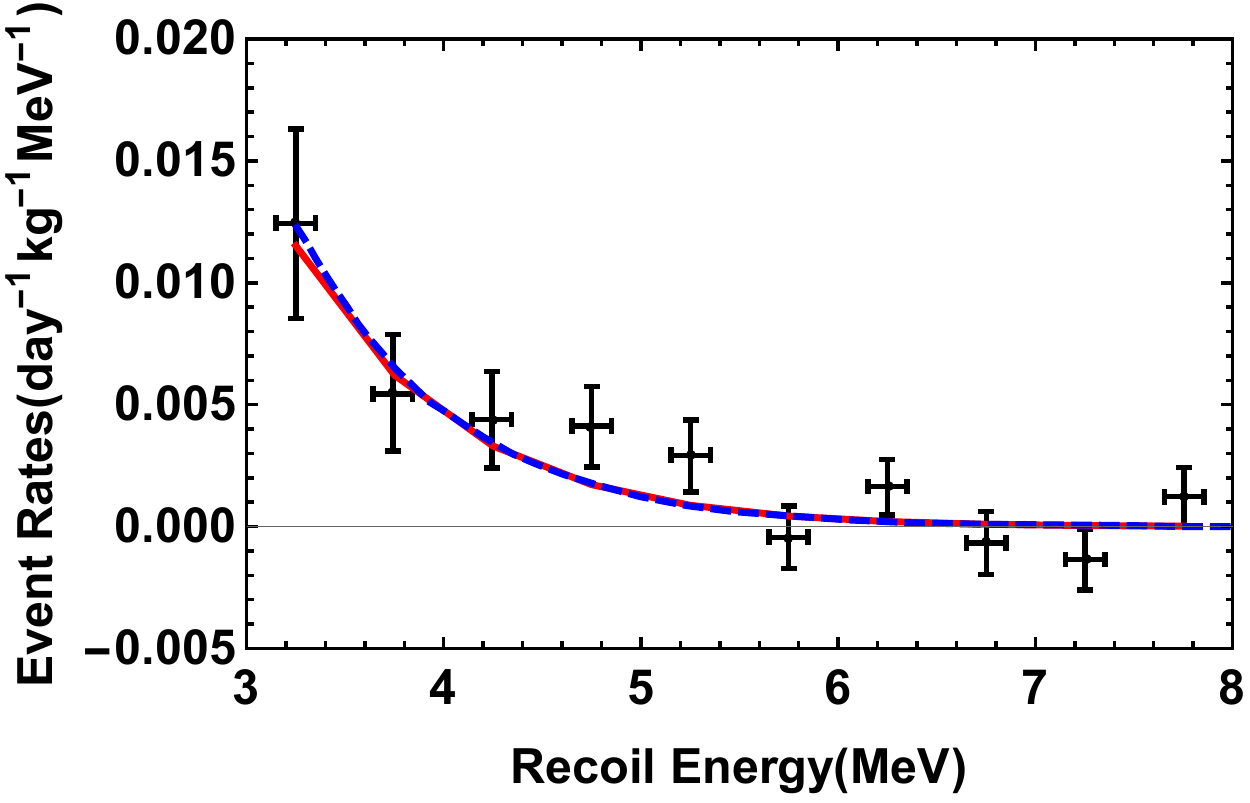}
	\end{center}
	\caption{The event rates of $\bar{\nu}_e-e^-$ scattering in the TEXONO experiment. The black data points and the blue dashed SM prediction curve are taken from Ref.~\cite{Deniz:2009mu}. The red curve corresponds to our best-fit prediction in the SM.
	}
	\label{fig:texono}
\end{figure}

The cross section of $\bar{\nu}_e$ scattering on electrons has been measured by the TEXONO experiment utilizing electron antineutrinos produced by the Kuo-Sheng Nuclear Power Reactor with a CsI(Tl) scintillating crystal detector~\cite{Deniz:2009mu}. The detector is placed at a distance of 28 m from the 2.9 GW reactor core. The range of recoil energy used in the analysis is from 3~MeV and 8~MeV, respectively. The measured event rates and uncertainties have been given in Ref.~\cite{Deniz:2009mu}, which are shown in Fig.~\ref{fig:texono}. As seen from the red and blue dashed curves in Fig.~\ref{fig:texono}, our SM predictions agree quite well with those given in the Ref.~\cite{Deniz:2009mu}. Therefore, we consider the following $\chi^2$ function in our analysis for new physics:
\begin{eqnarray}
\chi^2_\text{TEXONO}=\sum_i\frac{(R_i^0-R_i(1+\alpha))^2}{\sigma_{R,i}^2}+ \left(\frac{\alpha}{\sigma_\alpha}\right)^2\,,
\end{eqnarray}
where $R_i$ ($R_i^0$) and $\sigma_{R,i}$ are the predicted (measured) event rates and corresponding uncertainties in the $i$th recoil energy bin. Here $\sigma_\alpha$ is the normalization uncertainty, and we take it to be $5\%$ for conservation.
Both the measured event rates and uncertainties are taken from Ref.~\cite{Deniz:2009mu}. The predicted event rate in the $i$th recoil energy bin is calculated by
\begin{align}
R_i=N_e \int d E_\nu \phi_{\bar{\nu}_e}(E_\nu) \int_i dE_R \frac{d\sigma_{\bar{\nu}_e}}{dE_R}\eta(E_R)\,,
\end{align}
where $N_e=2.5\times10^{26}$ is the number of target electrons in the CsI detector per kilogram, $\phi_{\bar{\nu}_e}(E_\nu)$ is the reactor antineutrino flux, and $\eta(E_R)$ is the detection efficiency. Here we take it to be 100\%, which yields a good agreement for the SM predictions shown in Fig. 16 of Ref.~\cite{Deniz:2009mu}.

\subsection{Borexino}
The Borexino experiment measured solar neutrinos at the Laboratori Nazionali del Gran Sasso. Only $\nu_e$ are produced in the core of the Sun. However, after adiabatic propagation in the sun, the solar neutrinos arrive at the the Earth contain all three flavors: $\nu_e$, $\nu_\mu$ and $\nu_\tau$. The survival probability of solar neutrinos is given by~\cite{Liao:2017awz}
\begin{align}
P_{ee}\approx s_{13}^4+c_{13}^4(c_{12}^2\cos^2\theta_{12}^m+s_{12}^2\sin\theta_{12}^m)\,,
\label{eq:Pee}
\end{align}
where $s_{12}$ ($c_{12}$) denotes $\sin \theta_{12}$ ($\cos_{12}$) , $\theta_{12}$ is the vacuum mixing angles in the PMNS matrix. Here $\theta_{12}^m$ is the effective mixing angle at the production point in the Sun, which is given by
\begin{align}
\theta_{12}^m=\frac{1}{2}\arctan \frac{\sin2\theta_{12}}{\cos2\theta_{12}-\hat{A}c_{13}^2}\,,
\end{align}
with $\hat{A}\equiv2\sqrt{2}G_FN_e^SE_\nu/\delta m_{21}^2$ and $N_{e}^S$ being the number density of electron at the production point in the Sun. Here we ignore the small corrections due to the day-night asymmetry in the Borexino measurement~\cite{Bellini:2011yj}. We consider the \textit{pp}, $^7Be$ and \textit{pep} spectra measured in the Borexino phase-I~\cite{Bellini:2011rx, Collaboration:2011nga, Bellini:2013lnn} and phase-II~\cite{Bellini:2014uqa,Agostini:2017ixy}, and the $^8B$ data collected between January 2008 and December 2016~\cite{Agostini:2017cav}. The expected event rate at Borexino is given by
\begin{align}
R^i_\text{pre}=N_e \int d E_\nu \Phi^i(E_\nu) \left[P_{ee}^i\sigma_e(E_\nu)+(1-P_{ee}^i)\sigma_\mu(E_\nu)\right]\,,
\label{eq:Ri}
\end{align}
where $N_e=3.307\times10^{31}/ 100$ ton is the density of target electrons in the Borexino detector~\cite{Agostini:2017ixy}, and $i$ indicates solar neutrino sources \textit{pp}, $^7Be$, \textit{pep} and $^8B$. $\Phi^i(E_\nu)$ is the corresponding solar neutrino flux taken from the standard solar model (B16-GS98-HZ)~\cite{Vinyoles:2016djt}. $P_{ee}^i$ is the survival probability given in Eq.~(\ref{eq:Pee}). The cross section in Eq.~(\ref{eq:Ri}) is calculated by
\begin{align}
\sigma_\alpha=\int dE_R\frac{d\sigma_\alpha}{dE_R}\eta(E_R)\,,
\end{align}
where $\alpha=e,\mu$, $\frac{d\sigma_\alpha}{dE_R}$ is the differential cross section, and $\eta(E_R)$ is the detection efficiency. The detection efficiency for $^8B$ is extracted from Fig.~2 in Ref.~\cite{Agostini:2017cav} and we take $\eta(E_R)$ to be 100\% for other solar neutrino sources~\cite{Khan:2019jvr}. The measured event rates and our predicted event rates in the SM are given in Table~\ref{tab:borexino}. We see that our SM predictions agree with the measured event rates. For new physics analysis, we employ the following $\chi^2$ function~\cite{Khan:2019jvr}:
\begin{eqnarray}
\chi^2_\text{Borexino}=\sum_i\frac{\left[R^i_\text{exp}-R^i_\text{pre}(1+\alpha_i)\right]^2}{(\sigma^i_{stat})^2}+ \left(\frac{\alpha^i}{\sigma_\text{th}^i}\right)^2\,,
\end{eqnarray}
where $R^i_\text{exp}$ ($\sigma^i_{exp}$) are the central values (statistical uncertainties) of the $i$th measurement given in Table~\ref{tab:borexino}, $R^i_\text{pre}$ is the predicted event rates calculated in Eq.~(\ref{eq:Ri}), and $\sigma^i_{th}$ is the theoretical uncertainties given in the last column in Table~\ref{tab:borexino}.
\begin{table}
	\begin{center}
		\begin{tabular}{|c|c|c|c|}
			\hline
			Source &Measurement (cpd/100 t) & SM prediction (cpd/100 t)  & Percentage error \\
			\hline
			\textit{pp} & $134\pm10^{+6}_{-10} $ & $136.0\pm 1.6$ &  1.2\% \\
			\hline
			$^7Be$ & $46\pm1.5^{+1.5}_{-1.6}$(phase I), $48.3\pm1.1^{+0.4}_{-0.7}$(phase II) & $47.6\pm 2.9$ & 6.1\%\\
			\hline
			\textit{pep} & $3.1\pm0.6\pm 0.3$ (phase I), $2.43\pm0.36^{+0.15}_{-0.22}$ (phase II) & $2.76\pm 0.04$ & 1.3\%\\
			\hline
			$^8B$ & $0.223^{+0.015}_{-0.016} \pm 0.006$ & $0.209\pm 0.025$ & 12.0\%\\
			\hline
		\end{tabular}
	\end{center}
	\caption{The measured event rates at Borexino and our predicted event rates in the SM. The theoretical percentage uncertainties are given in the last column.
		\label{tab:borexino}}			
\end{table}

\section{Constraints from the experimental data}
%
In this section, we present our results of the constraints on the coupling coefficients of the general neutrino interactions with $\chi$ and electrons using the neutrino-electron scattering data from the CHARM-II, TEXONO and Borexino experiments.

\subsection{Flavor-universal bounds}
\begin{figure}[htb!]
	\begin{center}
		\includegraphics[scale=1,width=0.45\linewidth]{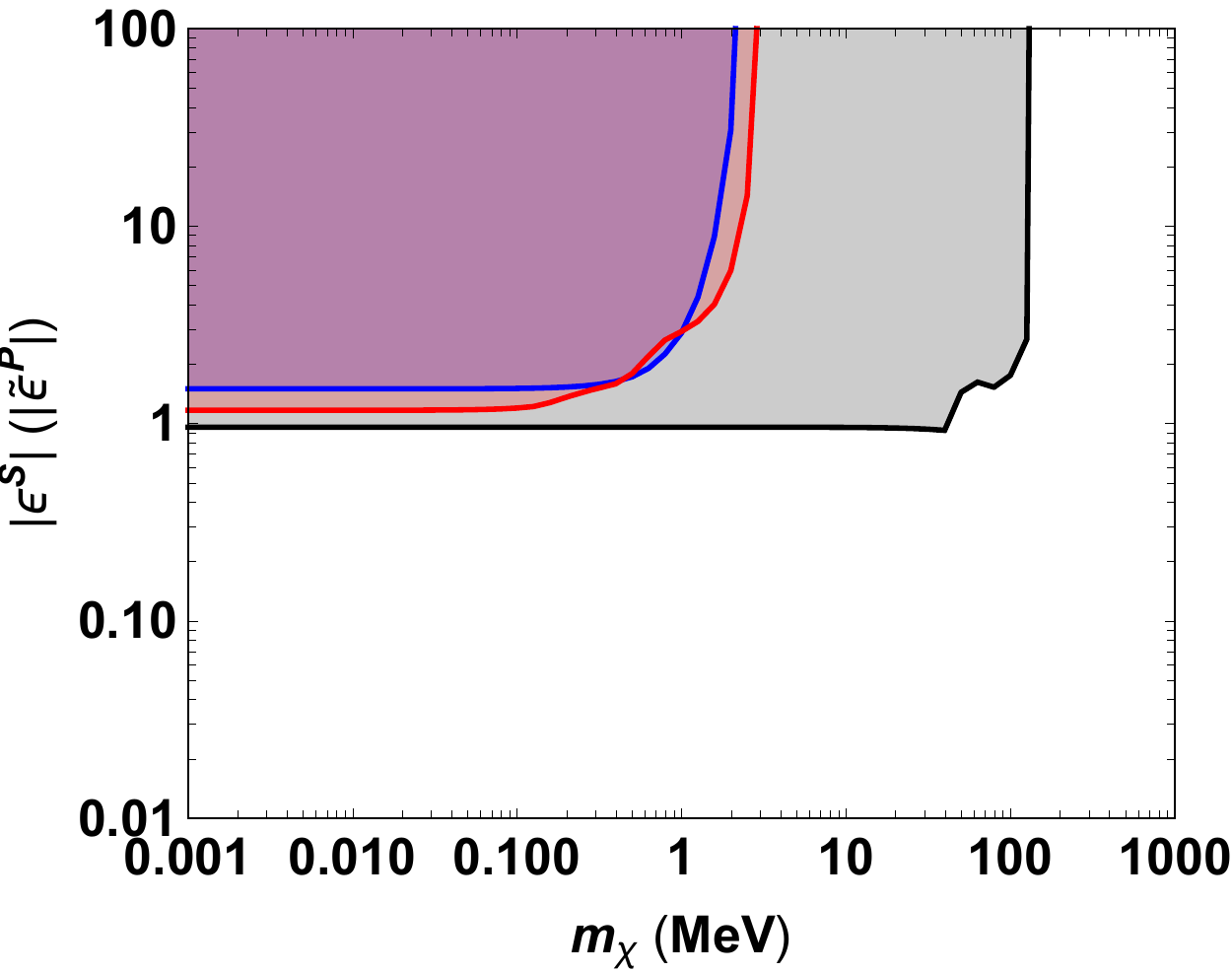}
		\includegraphics[scale=1,width=0.45\linewidth]{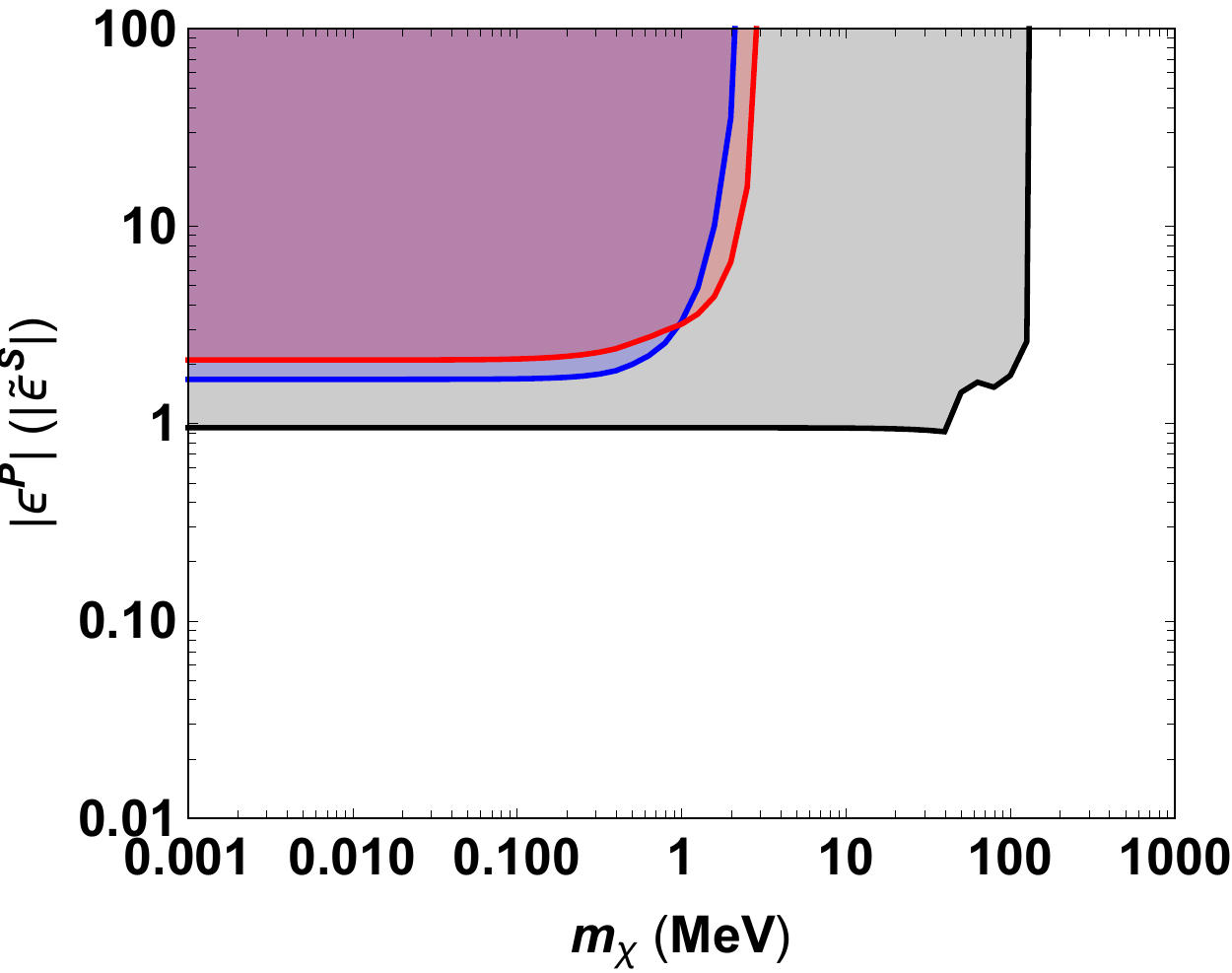}
		\includegraphics[scale=1,width=0.45\linewidth]{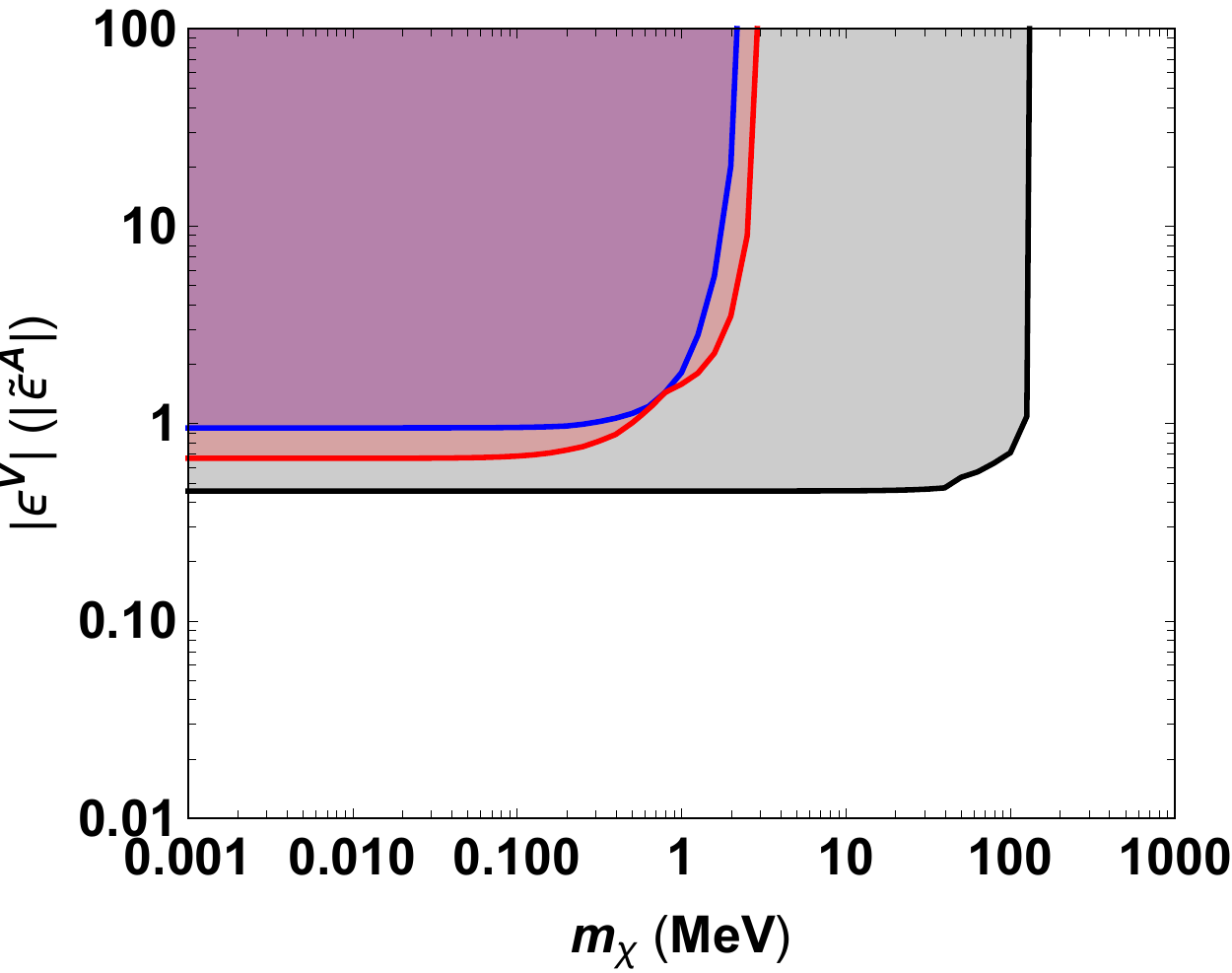}
		\includegraphics[scale=1,width=0.45\linewidth]{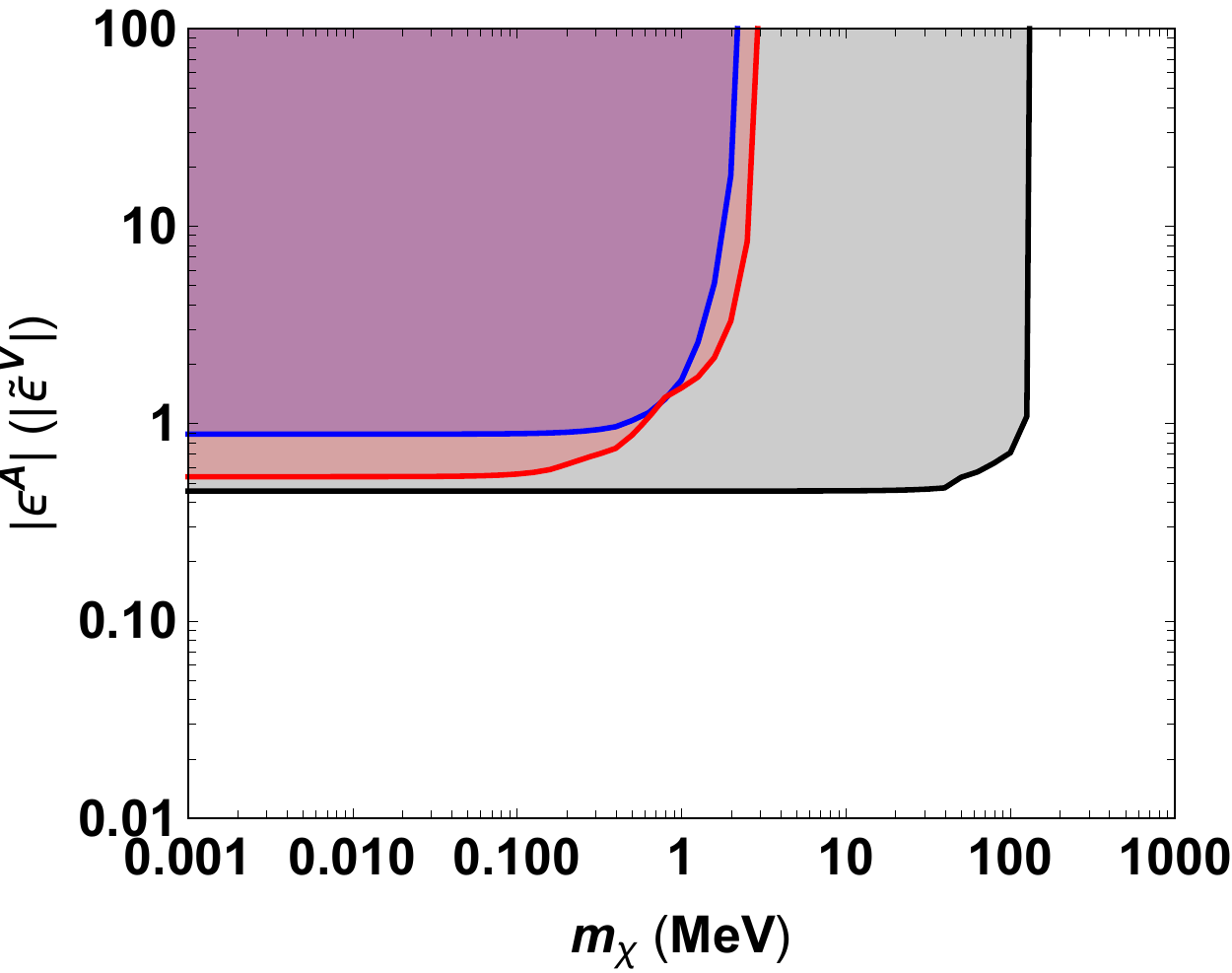}
		\includegraphics[scale=1,width=0.45\linewidth]{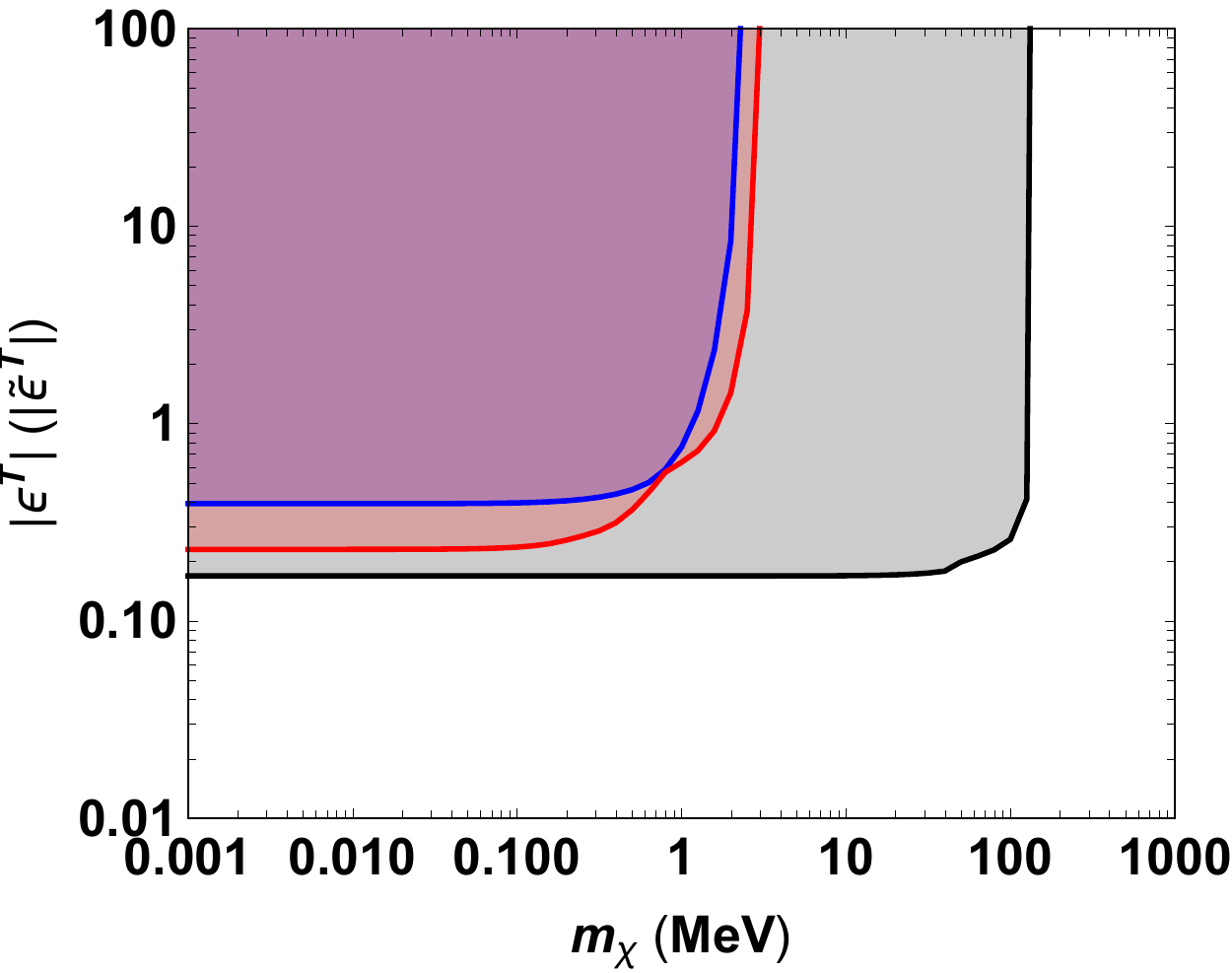}
		\includegraphics[scale=1,width=0.45\linewidth]{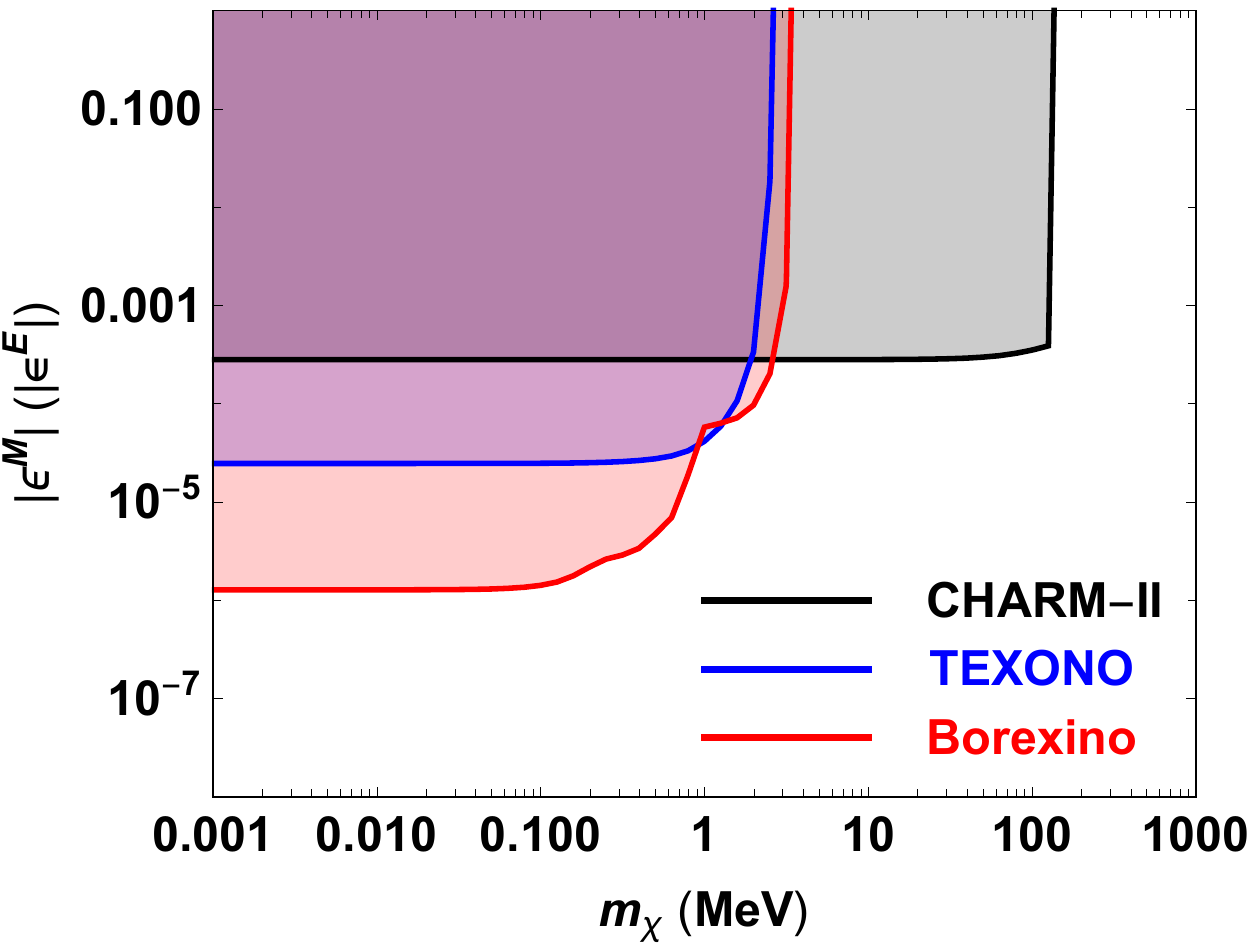}
	\end{center}
	\caption{The 90\% CL upper bounds on the magnitude of coupling coefficients as a function of $m_\chi$ for the  scalar, pseudoscalar, vector, axialvector, tensor and electromagnetic dipole interactions. We assume that the couplings are flavor universal and only one $\epsilon^i$ ($\tilde{\epsilon}^i$) exists at a time. The gray, blue and red shaded regions are excluded by the CHARM-II, TEXONO and Borexino experiment, respectively. The bounds on $|\tilde{\epsilon}_\alpha^S|, |\tilde{\epsilon}_\alpha^P|, |\tilde{\epsilon}_\alpha^V|, |\tilde{\epsilon}_\alpha^A|, |\tilde{\epsilon}_\alpha^T|$, and $|\epsilon_\alpha^{E}|$ are the same as $|\epsilon_\alpha^P|, |\epsilon_\alpha^S|, |\epsilon_\alpha^A|, |\epsilon_\alpha^V|, |\epsilon_\alpha^{T}|$, and $|\epsilon_\alpha^{M}|$, respectively.
	}
	\label{fig:bounds}
\end{figure}

We firstly consider the flavor-universal couplings, i.e. by setting $\epsilon_\alpha^i\equiv\epsilon^i$ in Eq.~(\ref{eq:Leff}), where $i$ indicates the scalar (S), pseudoscalar (P), vector(V), axialvector (A), tensor (T) and electromagnetic (E or M) dipole operators. Here we also assume only one $\epsilon^i$ ($\tilde{\epsilon}^i$) exists at a time. As mentioned before, the bounds on $\{\tilde{\epsilon}_\alpha^S, \tilde{\epsilon}_\alpha^P, \tilde{\epsilon}_\alpha^V, \tilde{\epsilon}_\alpha^A, \tilde{\epsilon}_\alpha^T, \epsilon_\alpha^{E}\}$ will be the same as $\{\epsilon_\alpha^P, \epsilon_\alpha^S, \epsilon_\alpha^A, \epsilon_\alpha^V, \epsilon_\alpha^{T}, \epsilon_\alpha^{M}\}$ in this case.
The 90\% CL upper bounds on the magnitude of the coefficients $\epsilon^i$ ($\tilde{\epsilon}^i$) as a function of $m_\chi$ are shown in Fig.~\ref{fig:bounds}. From Fig.~\ref{fig:bounds}, we see that the CHARM-II experiment yields the strongest bounds for the scalar, pseudoscalar, vector, axialvector and tensor interactions. For the electromagnetic dipole interaction, the Borexino experiment has the best sensitivity for $m_\chi$ below 1 MeV. There is an upper limit on $m_\chi$ for the CHARM-II bounds due to the kinematic constraint. This can be explained by Eq.~(\ref{eq:Enumin}), from which we get $m_\chi\leq\sqrt{(2E_\nu+m_e)m_e}-m_e\simeq 155$~MeV for $E_\nu=23.7$~GeV. Also, the upper limits on $m_\chi$ from the TEXONO and Borexino experiments are much smaller than the CHARM-II experiment due to low neutrino energies used in these two experiments. In addition, we see that the bounds become flat at small $m_\chi$, which can be understood from Eqs.~(\ref{eq:xsectionS}), (\ref{eq:xsectionP}), (\ref{eq:xsectionV}), (\ref{eq:xsectionA}), (\ref{eq:xsectionT}) and (\ref{eq:xsectionEM}) since the differential cross sections are insensitive to $m_\chi$ as $m_\chi\ll E_\nu$. From  Fig.~\ref{fig:bounds}, we find that for $m_\chi\lesssim 50$ MeV, the strongest bounds on the magnitude of $\epsilon^{S,P}$ ($\tilde{\epsilon}^{S,P}$), $\epsilon^{V,A}$ ($\tilde{\epsilon}^{V,A}$) and $\epsilon^T$  ($\tilde{\epsilon}^T$) can reach 1.0, 0.5 and 0.2, respectively. The strongest bounds on the magnitude of $\epsilon^{M,E}$ can reach
$1.3\times10^{-6}$ for $m_\chi\lesssim 0.1$ MeV.

\subsection{Flavor-dependent  bounds}
\begin{figure}[htb!]
	\begin{center}
		\includegraphics[scale=1,width=0.45\linewidth]{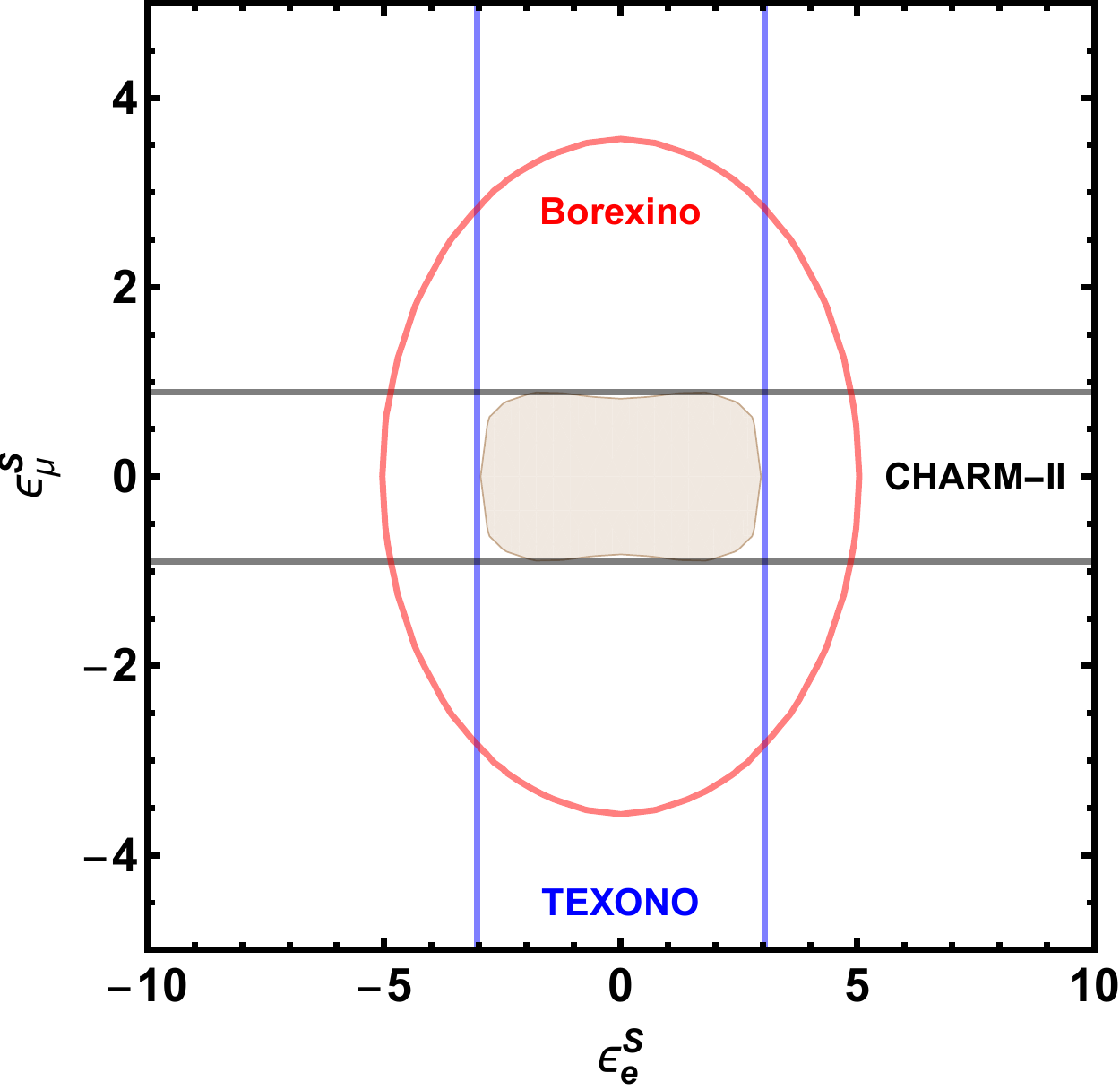}
		\includegraphics[scale=1,width=0.45\linewidth]{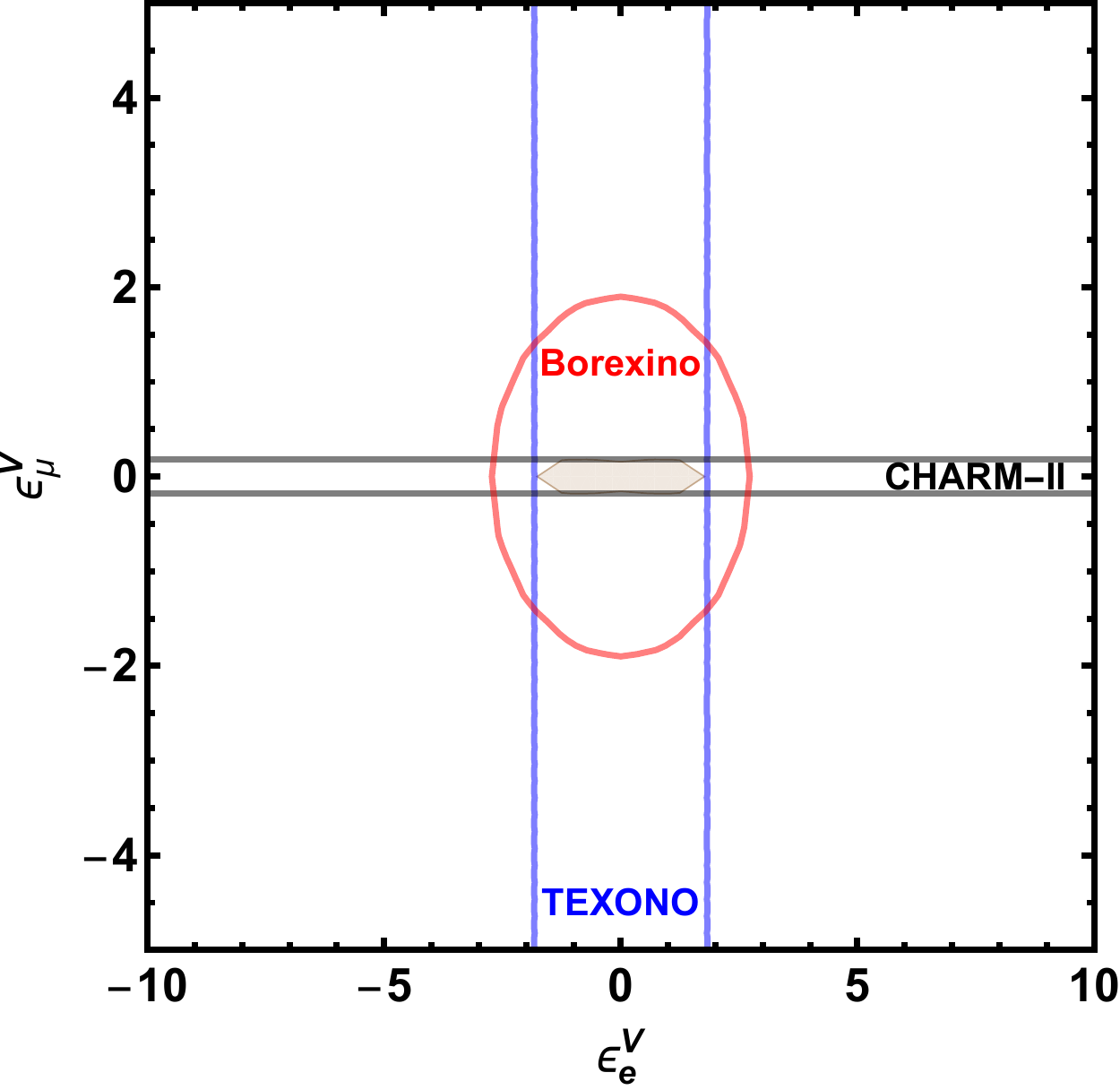}
	\end{center}
	\caption{The 90\% CL allowed regions in the  ($\epsilon_e$, $\epsilon_\mu$) plane for the scalar (left panel) and vector (right panel) interactions. Here we assume $m_\chi=1$ MeV and the coupling coefficients are real with $\epsilon_\tau=\epsilon_\mu$. The region enclosed by the black, blue and red curves correspond to the CHARM-II, TEXONO and Borexino experiments, respectively. The gray shaded regions correspond to the allowed regions of the combined data from these three experiments.
	}
	\label{fig:correlation}
\end{figure}

Since only $\nu_\mu$ ($\bar{\nu}_\mu$) are measured at the CHARM-II experiment and only $\bar{\nu}_e$ are measured at the TEXONO experiment, the bounds from these two experiments can be avoided if the coupling coefficients are flavor non-universal. To illustrate the flavor dependence of these bounds, we show the 90\% CL allowed regions in the  ($\epsilon_e$, $\epsilon_\mu$) plane for the scalar and vector interactions in Fig.~\ref{fig:correlation}. Here we fixed $m_\chi=1$ MeV, and assume the coupling coefficients are real for simplicity. We also assume $\epsilon_\tau=\epsilon_\mu$ for the Borexino experiments. As seen from Fig.~\ref{fig:correlation}, the CHARM-II experiment is not sensitive to $\epsilon_e$, and the TEXONO experiment has no sensitivity to $\epsilon_\mu$. The solar neutrino experiment at Borexino can impose constraints on both $\epsilon_e$ and $\epsilon_\mu$ due to the flavor transition in the Sun. We also show the allowed regions of the combined data from these three experiments as the gray shaded regions in Fig.~\ref{fig:correlation}. From Fig.~\ref{fig:correlation}, one can see that the sensitivity of the combined data mainly comes from the CHARM-II and TEXONO experiment.

\section{Prospects at DM experiments}
\label{sec:DM}

The exotic fermion $\chi$ can also serve as a DM particle if the correct DM abundance is satisfied. The relic abundance of DM is model dependent and changes based on the specific production mechanism and dynamics in the early Universe. The DM is either in thermal equilibrium with the SM bath and freezes out below some temperature, or is produced non-thermally. The thermal production of DM is a well-studied mechanism to obtain the correct abundance of DM in the early Universe. In our case, DM starts to keep in thermal equilibrium with the SM bath as a result of both the forward and backward processes $\bar{\chi}\chi\leftrightarrow \bar{\nu}\nu$ happening in the early Universe. As the Universe expands, the temperature drops below the DM mass and the DM falls out of thermal equilibrium. As a result, only the forward process of the annihilation of DM particles into neutrinos occurs and the abundance of DM is quickly suppressed so as to reach the current equilibrium abundance. This is the so-called freeze-out process for the production of DM. The DM in thermal equilibrium should not significantly modifies the standard predictions
of the ratios of light elemental abundances by the big bang nucleosynthesis (BBN). Also, the thermal DM species must not significantly alter the
temperature ratio of photons and neutrinos at recombination. It turns out that thermal DM lighter than MeV scale will contribute to the radiation energy density and be ruled out by $N_{eff}$ constraint~\cite{Boehm:2013jpa}. For the DM mass scale smaller than MeV, the above $\chi-\nu$ operators can induce the non-thermal production of DM after the SM neutrinos decouple and before the electrons leave the bath. In the case of non-thermal production, if the coupling to electron is large, the DM would suffer the constraint from overproduction near BBN. The calculation of the final DM abundance replies on the details of the UV completion. The annihilation cross section depends on the $\chi-\nu$ mixing and the nature of mediator such as a dark photon for vector interaction or a scalar for scalar interaction in a UV completion. We refer the UV models to Ref.~\cite{Dror:2020czw} and one can see that the XENON1T excess preferred parameter space can avoid overproducing the DM abundance.

Here we briefly discuss the detection of DM hypothesis and the XENON1T excess.
If we reverse the above process and interpret the exotic fermion $\chi$ as DM particle, the incoming DM $\chi$ can be absorbed by bound electron targets and emit a neutrino
\begin{eqnarray}
\chi e\to \nu e\;.
\end{eqnarray}
For the DM elastic scattering off the electron, to explain the XENON1T excess, the key point is how to produce abundant DM particles with high velocity $v_{\rm DM}\gtrsim 0.1$~\cite{Kannike:2020agf}. In contrast, Ref.~\cite{Dror:2020czw} proposed the above DM absorption scenario in which a DM particle deposits its mass energy rather than kinetic energy and it is sensitive to sub-MeV fermionic DM.
For the DM absorption with free electrons $\chi e\to \nu e$, analogous to the case with nucleus absorbing the DM~\cite{Dror:2019onn,Dror:2019dib}, the total event rate is naively given by
\begin{eqnarray}
R={\rho_\chi\over m_\chi} \sigma_e N_T \Theta(E_R^0-E_{\rm th})\;,
\label{rate}
\end{eqnarray}
where $N_T$ is the number of target nuclei per detector mass, the local DM density is $\rho_\chi\simeq 0.4~{\rm GeV}/{\rm cm}^3$, $\Theta$ is the Heaviside theta function, $E_{\rm th}$ is the experimental threshold, $E_R^0=m_\chi^2/2m_e$ for a free electron absorbing the DM, and $\sigma_e$ is the absorption cross section per electron. For the XENON1T experiment, we have $N_T\simeq 4\times 10^{27}/{\rm tonne}$ and $E_R^0\simeq 2$ keV giving $m_\chi\simeq 45$ keV. With the total exposure being 0.65 tonne$\cdot$years, XENON1T observed 285 events and the expected event number is $232\pm 15$. This gives the total scattering cross section $\sigma_e\simeq 2\times 10^{-48}~{\rm cm}^2$. In fact, the absorbing electron in a shell with a binding energy would be ionized with recoil energy~\cite{Essig:2011nj,Essig:2015cda}. In Eq.~(\ref{rate}) there should also appear the ionization form factor of an electron in a certain shell and the total differential ionization rate is obtained by summing over all possible shells of the absorbing target electrons. The recoil energy of a free electron absorbing the DM $E_R^0$ is then shifted. The best fit to the XENON1T data was found to be $(m_\chi=56.5~{\rm keV}, \sigma_e=1\times 10^{-49}~{\rm cm^2})$~\cite{Dror:2020czw}. We take the vector interaction for illustration. The total scattering cross section $\sigma_e$ for vector operators is
\begin{eqnarray}
\sigma_e^V &=& {G_F^2 m_\chi^2 (m_\chi + 2m_e)^2\over 32\pi (m_e+m_\chi)^4} \Big[|\epsilon^V|^2 (2m_e^2+4m_e m_\chi+3m_\chi^2)+|\tilde{\epsilon}^V|^2(6m_e^2+8m_e m_\chi+3m_\chi^2)\Big]\;.\nonumber \\
\end{eqnarray}
By transforming the above best fit to the parameterization in our context, one essentially obtains $|\epsilon^V|, |\tilde{\epsilon}^V|\simeq 0.1$ which is allowed by the neutrino scattering experiments. Note that the LUX-ZEPLIN (LZ) experiment at the Sanford Underground Research Facility (SURF) in South Dakota~\cite{Akerib:2019fml} will start to take data soon and is able to probe orders of magnitude more parameter space than XENON1T. Such next generation
detector will help to test the XENON1T favored region.

The decaying DM scenario usually faces the requirement of stability. The corresponding lifetime of $\chi$ should be longer than the age of the Universe, i.e. $t_{\rm Universe}=4.4\times 10^{17}~{\rm sec}$~\cite{Ade:2015xua}.
Requiring the DM being stable at the Universe time scale would set a very stringent bound on the coupling and/or the DM mass.
For the vector interaction in our effective framework, the leading decay process is $\chi\to \nu \gamma$ with one photon radiated from the closed electron loop.
The constraint would be quite stringent if only electron is involved in the calculation of the above decay width.
In a realistic UV model in Ref.~\cite{Dror:2020czw} with $\chi$ only coupled to right-handed neutrino, the decay $\chi\to \nu \gamma$ is suppressed by the insertion of neutrino mass. The decay of $\chi$ into $\nu\gamma\gamma$ is forbidden and the leading decay becomes $\chi\to \nu\gamma\gamma\gamma$ which leads to a quite weak constraint.
We refer the detailed discussion of the sub-MeV DM absorption by electrons to Ref.~\cite{Dror:2020czw} and future studies.
On the other hand, more stringent constraints can also be placed on decaying DM from the observations of the galactic and extra-galactic diffuse
X-ray or gamma-ray background. For several tens of keV DM, strong constraints have been obtained using INTEGRAL~\cite{Yuksel:2007xh,Boyarsky:2007ge,Essig:2013goa} and NuSTAR~\cite{Perez:2016tcq,Roach:2019ctw}. The bound on the lifetime of decaying DM depends on the specific model parameters and decay topologies.

\section{Conclusion}
\label{sec:Con}

In this work we study the constraints on general neutrino interactions with sub-GeV exotic fermion $\chi$ from neutrino-electron scattering experiments.
The general neutrino interactions are composed of dimension-5 dipole operators and dimension-6 four-fermion operators.
We employ the measurements of CHARM-II, TEXONO and Borexino experiments to set limits on the neutrino-electron scattering with an outgoing fermion $\chi$. We find that the bounds are dominated by the CHARM-II experiment in most of the parameter space for the flavor-universal interactions and $m_\chi$ below 155 MeV, while the Borexino experiment sets the strongest bounds in the low mass region for the electromagnetic dipole interactions. The limits are found to be $|\epsilon^{S,P}|(|\tilde{\epsilon}^{S,P}|)<1$, $|\epsilon^{V,A}|(|\tilde{\epsilon}^{V,A}|)<0.5$, $|\epsilon^{T}|(|\tilde{\epsilon}^{T}|)<0.2$ for $m_\chi\lesssim 50$ MeV and $|\epsilon^{M,E}|<1.3\times 10^{-6}$ for $m_\chi\lesssim 0.1$ MeV. If the coupling coefficients are flavor non-universal, the bounds on $\epsilon_e$ ($\epsilon_\mu$) can be avoided for the CHARM-II (TEXONO) experiment, and there are correlations between the bounds on the coupling coefficients from the Borexino experiment. Finally, as an example, we discuss the detection of sub-MeV DM absorbed by bound electron targets. By transforming the best fit to the XENON1T data in our parameterization, we obtain the preferred coefficients for vector interactions as $|\epsilon^V|, |\tilde{\epsilon}^V|\simeq 0.1$ which is allowed by the neutrino experiments.

\acknowledgments
TL is supported by the National Natural Science Foundation of China (Grant No. 12035008, 11975129) and ``the Fundamental Research Funds for the Central Universities'', Nankai University (Grant No. 63196013). JL is supported by the National Natural Science Foundation of China (Grant No. 11905299), Guangdong Basic and Applied Basic Research Foundation (Grant No. 2020A1515011479), the  Fundamental  Research  Funds  for  the  Central Universities, and the Sun Yat-Sen University Science Foundation.

\appendix

\section{Calculation of the differential cross section}
\label{app:diffcalc}
The amplitude for $\nu(p_1) e(k_1) \to \chi(p_2) e(k_2)$ is given by
\begin{eqnarray}
\mathcal{M}&=&{G_F\over \sqrt{2}} \bar{u}_\chi(p_2) P_L u_\nu(p_1)~\bar{u}_e(k_2) (\epsilon^S+\tilde{\epsilon}^S \gamma_5) u_e(k_1) \nonumber \\
&+& {G_F\over \sqrt{2}} \bar{u}_\chi(p_2) i \gamma_5 P_L u_\nu(p_1)~\bar{u}_e(k_2) i\gamma_5(\epsilon^P+\tilde{\epsilon}^P \gamma_5) u_e(k_1)\nonumber \\
&+&{G_F\over \sqrt{2}} \bar{u}_\chi(p_2) \gamma_\mu P_L u_\nu(p_1)~\bar{u}_e(k_2) \gamma^\mu(\epsilon^V+\tilde{\epsilon}^V \gamma_5) u_e(k_1)\nonumber \\
&+& {G_F\over \sqrt{2}} \bar{u}_\chi(p_2) \gamma_\mu \gamma_5 P_L u_\nu(p_1)~\bar{u}_e(k_2) \gamma^\mu \gamma_5(\epsilon^A+\tilde{\epsilon}^A \gamma_5) u_e(k_1)\nonumber \\
&+& {G_F\over \sqrt{2}} \bar{u}_\chi(p_2) \sigma_{\mu\nu} P_L u_\nu(p_1)~\bar{u}_e(k_2) \sigma^{\mu\nu}(\epsilon^T+\tilde{\epsilon}^T \gamma_5) u_e(k_1)\nonumber \\
&+&{i G_Fv_He Q_e\over t^2}\bar{u}_\chi(p_2) \sigma_{\mu\nu} (\epsilon^M+\epsilon^E\gamma_5) P_L u_\nu(p_1)~\bar{u}_e(k_2)\gamma_\mu t_\nu u_e(k_1)\;,
\end{eqnarray}
where the projector $P_L=(1-\gamma_5)/2$ is inserted to force the incoming neutrinos to be left-handed and $t=p_1-p_2$.
The amplitude for $\overline{\nu}(p_1) e(k_1) \to \overline{\chi}(p_2) e(k_2)$ is given by
\begin{eqnarray}
\mathcal{M}&=&{G_F\over \sqrt{2}} \bar{v}_\nu(p_1) P_R v_\chi(p_2)~\bar{u}_e(k_2) (\epsilon^{S\ast}-\tilde{\epsilon}^{S\ast} \gamma_5) u_e(k_1) \nonumber \\
&+& {G_F\over \sqrt{2}} \bar{v}_\nu(p_1) P_R i \gamma_5 v_\chi(p_2)~\bar{u}_e(k_2) i\gamma_5(\epsilon^{P\ast}-\tilde{\epsilon}^{P\ast} \gamma_5) u_e(k_1)\nonumber \\
&+& {G_F\over \sqrt{2}} \bar{v}_\nu(p_1) P_R \gamma_\mu v_\chi(p_2)~\bar{u}_e(k_2) \gamma^\mu(\epsilon^{V\ast}+\tilde{\epsilon}^{V\ast} \gamma_5) u_e(k_1)\nonumber \\
&+& {G_F\over \sqrt{2}} \bar{v}_\nu(p_1) P_R \gamma_\mu \gamma_5 v_\chi(p_2)~\bar{u}_e(k_2) \gamma^\mu \gamma_5(\epsilon^{A\ast}+\tilde{\epsilon}^{A\ast} \gamma_5) u_e(k_1)\nonumber \\
&+& {G_F\over \sqrt{2}} \bar{v}_\nu(p_1) P_R \sigma_{\mu\nu} v_\chi(p_2)~\bar{u}_e(k_2) \sigma^{\mu\nu}(\epsilon^{T\ast}-\tilde{\epsilon}^{T\ast} \gamma_5) u_e(k_1)\nonumber \\
&+&{iG_Fv_He Q_e\over t^2} \bar{v}_\nu(p_1) P_R \sigma_{\mu\nu} (\epsilon^{M\ast}-\epsilon^{E\ast}\gamma_5)v_\chi(p_2)~\bar{u}_e(k_2)\gamma_\mu t_\nu u_e(k_1)\;,
\end{eqnarray}
where the projector $P_R=(1+\gamma_5)/2$ is inserted to force the incoming anti-neutrinos to be right-handed.
The differential cross section of neutrino-electron scattering $\nu(\bar{\nu}) + e\to \chi (\bar{\chi}) + e$ is
\begin{eqnarray}
{d\sigma(\nu e)\over dE_R}={1\over 32\pi m_e E_\nu^2} \overline{|\mathcal{M}|^2}\;,
\label{eq:dsigmadT}
\end{eqnarray}
where $\overline{|\mathcal{M}|^2}$ is the spin-averaged amplitude square.
The scattering angle is
\begin{eqnarray}
\cos\theta = {E_R(E_\nu + m_e)+m_\chi^2/2\over E_\nu \sqrt{E_R^2+2m_e E_R}}\;.
\end{eqnarray}
By requiring $\cos\theta\leq1$, we can get the bounds on $E_R$ as
\begin{eqnarray}
E_R^{\rm min(max)} = {2m_e E_\nu^2-m_\chi^2(m_e+E_\nu)\mp E_\nu \sqrt{(2m_e E_\nu-m_\chi^2)^2-4m_e^2 m_\chi^2}\over 2m_e(m_e + 2E_\nu)}\;,
\end{eqnarray}
and the minimal energy to generate the elastic scattering is
\begin{eqnarray}
E_\nu^{\rm min} = m_\chi+{m_\chi^2 \over 2 m_e}\;.
\label{eq:Enumin}
\end{eqnarray}

\bibliographystyle{JHEP}
\bibliography{refs}

\providecommand{\href}[2]{#2}\begingroup\raggedright\begin{thebibliography}{10}

\bibitem{Zyla:2020zbs}
{\scshape Particle Data Group} collaboration, P.~A. Zyla et~al., \emph{{Review
  of Particle Physics}},
  \href{http://dx.doi.org/10.1093/ptep/ptaa104}{\emph{PTEP} {\bf 2020} (2020)
  083C01}.

\bibitem{Brdar:2018qqj}
V.~Brdar, W.~Rodejohann and X.-J. Xu, \emph{{Producing a new Fermion in
  Coherent Elastic Neutrino-Nucleus Scattering: from Neutrino Mass to Dark
  Matter}}, \href{http://dx.doi.org/10.1007/JHEP12(2018)024}{\emph{JHEP} {\bf
  12} (2018) 024}, [\href{http://arxiv.org/abs/1810.03626}{{\tt 1810.03626}}].

\bibitem{Chang:2019sel}
W.-F. Chang and J.~N. Ng, \emph{{KeV scale new fermion from a hidden sector}},
  \href{http://dx.doi.org/10.1103/PhysRevD.101.035028}{\emph{Phys. Rev. D} {\bf
  101} (2020) 035028}, [\href{http://arxiv.org/abs/1903.12545}{{\tt
  1903.12545}}].

\bibitem{Dror:2019onn}
J.~A. Dror, G.~Elor and R.~Mcgehee, \emph{{Directly Detecting Signals from
  Absorption of Fermionic Dark Matter}},
  \href{http://dx.doi.org/10.1103/PhysRevLett.124.181301}{\emph{Phys. Rev.
  Lett.} {\bf 124} (2020) 18}, [\href{http://arxiv.org/abs/1905.12635}{{\tt
  1905.12635}}].

\bibitem{Dror:2019dib}
J.~A. Dror, G.~Elor and R.~Mcgehee, \emph{{Absorption of Fermionic Dark Matter
  by Nuclear Targets}},
  \href{http://dx.doi.org/10.1007/JHEP02(2020)134}{\emph{JHEP} {\bf 02} (2020)
  134}, [\href{http://arxiv.org/abs/1908.10861}{{\tt 1908.10861}}].

\bibitem{Chang:2020jwl}
W.-F. Chang and J.~Liao, \emph{{Constraints on light singlet fermion
  interactions from coherent elastic neutrino-nucleus scattering}},
  \href{http://dx.doi.org/10.1103/PhysRevD.102.075004}{\emph{Phys. Rev. D} {\bf
  102} (2020) 075004}, [\href{http://arxiv.org/abs/2002.10275}{{\tt
  2002.10275}}].

\bibitem{Li:2020lba}
T.~Li, X.-D. Ma and M.~A. Schmidt, \emph{{General neutrino interactions with
  sterile neutrinos in light of coherent neutrino-nucleus scattering and meson
  invisible decays}},
  \href{http://dx.doi.org/10.1007/JHEP07(2020)152}{\emph{JHEP} {\bf 07} (2020)
  152}, [\href{http://arxiv.org/abs/2005.01543}{{\tt 2005.01543}}].

\bibitem{Hurtado:2020vlj}
N.~Hurtado, H.~Mir, I.~M. Shoemaker, E.~Welch and J.~Wyenberg, \emph{{Dark
  Matter-Neutrino Interconversion at COHERENT, Direct Detection, and the Early
  Universe}}, \href{http://dx.doi.org/10.1103/PhysRevD.102.015006}{\emph{Phys.
  Rev. D} {\bf 102} (2020) 015006},
  [\href{http://arxiv.org/abs/2005.13384}{{\tt 2005.13384}}].

\bibitem{Li:2020pfy}
T.~Li and J.~Liao, \emph{{Loop effect in the coherent neutrino-nucleus
  scattering}}, \href{http://dx.doi.org/10.1007/JHEP02(2021)099}{\emph{JHEP}
  {\bf 02} (2021) 099}, [\href{http://arxiv.org/abs/2008.00743}{{\tt
  2008.00743}}].

\bibitem{Ge:2020jfn}
S.-F. Ge, P.~Pasquini and J.~Sheng, \emph{{Solar neutrino scattering with
  electron into massive sterile neutrino}},
  \href{http://dx.doi.org/10.1016/j.physletb.2020.135787}{\emph{Phys. Lett. B}
  {\bf 810} (2020) 135787}, [\href{http://arxiv.org/abs/2006.16069}{{\tt
  2006.16069}}].

\bibitem{Shoemaker:2020kji}
I.~M. Shoemaker, Y.-D. Tsai and J.~Wyenberg, \emph{{An Active-to-Sterile
  Neutrino Transition Dipole Moment and the XENON1T Excess}},
  \href{http://arxiv.org/abs/2007.05513}{{\tt 2007.05513}}.

\bibitem{Shakeri:2020wvk}
S.~Shakeri, F.~Hajkarim and S.-S. Xue, \emph{{Shedding New Light on Sterile
  Neutrinos from XENON1T Experiment}},
  \href{http://arxiv.org/abs/2008.05029}{{\tt 2008.05029}}.

\bibitem{Dror:2020czw}
J.~A. Dror, G.~Elor, R.~McGehee and T.-T. Yu, \emph{{Absorption of sub-MeV
  fermionic dark matter by electron targets}},
  \href{http://dx.doi.org/10.1103/PhysRevD.103.035001}{\emph{Phys. Rev. D} {\bf
  103} (2021) 035001}, [\href{http://arxiv.org/abs/2011.01940}{{\tt
  2011.01940}}].

\bibitem{Brdar:2020quo}
V.~Brdar, A.~Greljo, J.~Kopp and T.~Opferkuch, \emph{{The Neutrino Magnetic
  Moment Portal: Cosmology, Astrophysics, and Direct Detection}},
  \href{http://dx.doi.org/10.1088/1475-7516/2021/01/039}{\emph{JCAP} {\bf 01}
  (2021) 039}, [\href{http://arxiv.org/abs/2007.15563}{{\tt 2007.15563}}].

\bibitem{AristizabalSierra:2020zod}
D.~Aristizabal~Sierra, R.~Branada, O.~G. Miranda and G.~Sanchez~Garcia,
  \emph{{Sensitivity of direct detection experiments to neutrino magnetic
  dipole moments}},
  \href{http://dx.doi.org/10.1007/JHEP12(2020)178}{\emph{JHEP} {\bf 12} (2020)
  178}, [\href{http://arxiv.org/abs/2008.05080}{{\tt 2008.05080}}].

\bibitem{Aprile:2020tmw}
{\scshape XENON} collaboration, E.~Aprile et~al., \emph{{Excess electronic
  recoil events in XENON1T}},
  \href{http://dx.doi.org/10.1103/PhysRevD.102.072004}{\emph{Phys. Rev. D} {\bf
  102} (2020) 072004}, [\href{http://arxiv.org/abs/2006.09721}{{\tt
  2006.09721}}].

\bibitem{Viaux:2013lha}
N.~Viaux, M.~Catelan, P.~B. Stetson, G.~Raffelt, J.~Redondo, A.~A.~R. Valcarce
  et~al., \emph{{Neutrino and axion bounds from the globular cluster M5 (NGC
  5904)}}, \href{http://dx.doi.org/10.1103/PhysRevLett.111.231301}{\emph{Phys.
  Rev. Lett.} {\bf 111} (2013) 231301},
  [\href{http://arxiv.org/abs/1311.1669}{{\tt 1311.1669}}].

\bibitem{Diaz:2019kim}
S.~A. D\'\i{}az, K.-P. Schr\"oder, K.~Zuber, D.~Jack and E.~E.~B. Barrios,
  \emph{{Constraint on the axion-electron coupling constant and the neutrino
  magnetic dipole moment by using the tip-RGB luminosity of fifty globular
  clusters}},  \href{http://arxiv.org/abs/1910.10568}{{\tt 1910.10568}}.

\bibitem{DiLuzio:2020jjp}
L.~Di~Luzio, M.~Fedele, M.~Giannotti, F.~Mescia and E.~Nardi, \emph{{Solar
  axions cannot explain the XENON1T excess}},
  \href{http://dx.doi.org/10.1103/PhysRevLett.125.131804}{\emph{Phys. Rev.
  Lett.} {\bf 125} (2020) 131804}, [\href{http://arxiv.org/abs/2006.12487}{{\tt
  2006.12487}}].

\bibitem{Gao:2020wer}
C.~Gao, J.~Liu, L.-T. Wang, X.-P. Wang, W.~Xue and Y.-M. Zhong,
  \emph{{Reexamining the Solar Axion Explanation for the XENON1T Excess}},
  \href{http://dx.doi.org/10.1103/PhysRevLett.125.131806}{\emph{Phys. Rev.
  Lett.} {\bf 125} (2020) 131806}, [\href{http://arxiv.org/abs/2006.14598}{{\tt
  2006.14598}}].

\bibitem{Dent:2020jhf}
J.~B. Dent, B.~Dutta, J.~L. Newstead and A.~Thompson, \emph{{Inverse Primakoff
  Scattering as a Probe of Solar Axions at Liquid Xenon Direct Detection
  Experiments}},
  \href{http://dx.doi.org/10.1103/PhysRevLett.125.131805}{\emph{Phys. Rev.
  Lett.} {\bf 125} (2020) 131805}, [\href{http://arxiv.org/abs/2006.15118}{{\tt
  2006.15118}}].

\bibitem{Agostini:2017ixy}
{\scshape Borexino} collaboration, M.~Agostini et~al., \emph{{First
  Simultaneous Precision Spectroscopy of $pp$, $^7$Be, and $pep$ Solar
  Neutrinos with Borexino Phase-II}},
  \href{http://dx.doi.org/10.1103/PhysRevD.100.082004}{\emph{Phys. Rev. D} {\bf
  100} (2019) 082004}, [\href{http://arxiv.org/abs/1707.09279}{{\tt
  1707.09279}}].

\bibitem{Vilain:1993kd}
{\scshape CHARM-II} collaboration, P.~Vilain et~al., \emph{{Measurement of
  differential cross-sections for muon-neutrino electron scattering}},
  \href{http://dx.doi.org/10.1016/0370-2693(93)90408-A}{\emph{Phys. Lett. B}
  {\bf 302} (1993) 351--355}.

\bibitem{Vilain:1994qy}
{\scshape CHARM-II} collaboration, P.~Vilain et~al., \emph{{Precision
  measurement of electroweak parameters from the scattering of muon-neutrinos
  on electrons}},
  \href{http://dx.doi.org/10.1016/0370-2693(94)91421-4}{\emph{Phys. Lett. B}
  {\bf 335} (1994) 246--252}.

\bibitem{Deniz:2009mu}
{\scshape TEXONO} collaboration, M.~Deniz et~al., \emph{{Measurement of
  Nu(e)-bar -Electron Scattering Cross-Section with a CsI(Tl) Scintillating
  Crystal Array at the Kuo-Sheng Nuclear Power Reactor}},
  \href{http://dx.doi.org/10.1103/PhysRevD.81.072001}{\emph{Phys. Rev. D} {\bf
  81} (2010) 072001}, [\href{http://arxiv.org/abs/0911.1597}{{\tt 0911.1597}}].

\bibitem{Khan:2014zwa}
A.~N. Khan, D.~W. McKay and F.~Tahir, \emph{{Short baseline reactor
  $\bar{\nu}-e$ scattering experiments and nonstandard neutrino interactions at
  source and detector}},
  \href{http://dx.doi.org/10.1103/PhysRevD.90.053008}{\emph{Phys. Rev. D} {\bf
  90} (2014) 053008}, [\href{http://arxiv.org/abs/1407.4263}{{\tt 1407.4263}}].

\bibitem{Khan:2016uon}
A.~N. Khan, \emph{{Global analysis of the source and detector nonstandard
  interactions using the short baseline \ensuremath{\nu}-e and
  \ensuremath{\nu}\textasciimacron{}-e scattering data}},
  \href{http://dx.doi.org/10.1103/PhysRevD.93.093019}{\emph{Phys. Rev. D} {\bf
  93} (2016) 093019}, [\href{http://arxiv.org/abs/1605.09284}{{\tt
  1605.09284}}].

\bibitem{Khan:2017oxw}
A.~N. Khan and D.~W. McKay, \emph{{$\sin^2(\theta)w$ estimate and bounds on
  nonstandard interactions at source and detector in the solar neutrino
  low-energy regime}},
  \href{http://dx.doi.org/10.1007/JHEP07(2017)143}{\emph{JHEP} {\bf 07} (2017)
  143}, [\href{http://arxiv.org/abs/1704.06222}{{\tt 1704.06222}}].

\bibitem{Khan:2017djo}
A.~N. Khan, \emph{{$\sin ^{2}\theta _{W}$ Estimate and Neutrino Electromagnetic
  Properties from Low-Energy Solar Data}},
  \href{http://dx.doi.org/10.1088/1361-6471/ab0057}{\emph{J. Phys. G} {\bf 46}
  (2019) 035005}, [\href{http://arxiv.org/abs/1709.02930}{{\tt 1709.02930}}].

\bibitem{Lindner:2018kjo}
M.~Lindner, F.~S. Queiroz, W.~Rodejohann and X.-J. Xu, \emph{{Neutrino-electron
  scattering: general constraints on Z$^{'}$ and dark photon models}},
  \href{http://dx.doi.org/10.1007/JHEP05(2018)098}{\emph{JHEP} {\bf 05} (2018)
  098}, [\href{http://arxiv.org/abs/1803.00060}{{\tt 1803.00060}}].

\bibitem{Khan:2019jvr}
A.~N. Khan, W.~Rodejohann and X.-J. Xu, \emph{{Borexino and general neutrino
  interactions}},
  \href{http://dx.doi.org/10.1103/PhysRevD.101.055047}{\emph{Phys. Rev. D} {\bf
  101} (2020) 055047}, [\href{http://arxiv.org/abs/1906.12102}{{\tt
  1906.12102}}].

\bibitem{Jodlowski:2020vhr}
Lowski and S.~Trojanowski, \emph{{Neutrino beam-dump experiment with FASER at
  the LHC}},  \href{http://arxiv.org/abs/2011.04751}{{\tt 2011.04751}}.

\bibitem{Batell:2021blf}
B.~Batell, J.~L. Feng and S.~Trojanowski, \emph{{Detecting Dark Matter with
  Far-Forward Emulsion and Liquid Argon Detectors at the LHC}},
  \href{http://arxiv.org/abs/2101.10338}{{\tt 2101.10338}}.

\bibitem{Allen:1992qe}
R.~C. Allen et~al., \emph{{Study of electron-neutrino electron elastic
  scattering at LAMPF}},
  \href{http://dx.doi.org/10.1103/PhysRevD.47.11}{\emph{Phys. Rev. D} {\bf 47}
  (1993) 11--28}.

\bibitem{Auerbach:2001wg}
{\scshape LSND} collaboration, L.~B. Auerbach et~al., \emph{{Measurement of
  electron - neutrino - electron elastic scattering}},
  \href{http://dx.doi.org/10.1103/PhysRevD.63.112001}{\emph{Phys. Rev. D} {\bf
  63} (2001) 112001}, [\href{http://arxiv.org/abs/hep-ex/0101039}{{\tt
  hep-ex/0101039}}].

\bibitem{Park:2015eqa}
{\scshape MINERvA} collaboration, J.~Park et~al., \emph{{Measurement of
  Neutrino Flux from Neutrino-Electron Elastic Scattering}},
  \href{http://dx.doi.org/10.1103/PhysRevD.93.112007}{\emph{Phys. Rev. D} {\bf
  93} (2016) 112007}, [\href{http://arxiv.org/abs/1512.07699}{{\tt
  1512.07699}}].

\bibitem{Valencia:2019mkf}
{\scshape MINERvA} collaboration, E.~Valencia et~al., \emph{{Constraint of the
  MINER$\nu$A medium energy neutrino flux using neutrino-electron elastic
  scattering}},
  \href{http://dx.doi.org/10.1103/PhysRevD.100.092001}{\emph{Phys. Rev. D} {\bf
  100} (2019) 092001}, [\href{http://arxiv.org/abs/1906.00111}{{\tt
  1906.00111}}].

\bibitem{Rodejohann:2017vup}
W.~Rodejohann, X.-J. Xu and C.~E. Yaguna, \emph{{Distinguishing between Dirac
  and Majorana neutrinos in the presence of general interactions}},
  \href{http://dx.doi.org/10.1007/JHEP05(2017)024}{\emph{JHEP} {\bf 05} (2017)
  024}, [\href{http://arxiv.org/abs/1702.05721}{{\tt 1702.05721}}].

\bibitem{Arguelles:2018mtc}
C.~A. Arg\"uelles, M.~Hostert and Y.-D. Tsai, \emph{{Testing New Physics
  Explanations of the MiniBooNE Anomaly at Neutrino Scattering Experiments}},
  \href{http://dx.doi.org/10.1103/PhysRevLett.123.261801}{\emph{Phys. Rev.
  Lett.} {\bf 123} (2019) 261801}, [\href{http://arxiv.org/abs/1812.08768}{{\tt
  1812.08768}}].

\bibitem{Liao:2017awz}
J.~Liao, D.~Marfatia and K.~Whisnant, \emph{{Nonstandard interactions in solar
  neutrino oscillations with Hyper-Kamiokande and JUNO}},
  \href{http://dx.doi.org/10.1016/j.physletb.2017.05.054}{\emph{Phys. Lett. B}
  {\bf 771} (2017) 247--253}, [\href{http://arxiv.org/abs/1704.04711}{{\tt
  1704.04711}}].

\bibitem{Bellini:2011yj}
{\scshape Borexino} collaboration, G.~Bellini et~al., \emph{{Absence of
  day--night asymmetry of 862 keV $^7$Be solar neutrino rate in Borexino and
  MSW oscillation parameters}},
  \href{http://dx.doi.org/10.1016/j.physletb.2011.11.025}{\emph{Phys. Lett. B}
  {\bf 707} (2012) 22--26}, [\href{http://arxiv.org/abs/1104.2150}{{\tt
  1104.2150}}].

\bibitem{Bellini:2011rx}
G.~Bellini et~al., \emph{{Precision measurement of the 7Be solar neutrino
  interaction rate in Borexino}},
  \href{http://dx.doi.org/10.1103/PhysRevLett.107.141302}{\emph{Phys. Rev.
  Lett.} {\bf 107} (2011) 141302}, [\href{http://arxiv.org/abs/1104.1816}{{\tt
  1104.1816}}].

\bibitem{Collaboration:2011nga}
{\scshape Borexino} collaboration, G.~Bellini et~al., \emph{{First evidence of
  pep solar neutrinos by direct detection in Borexino}},
  \href{http://dx.doi.org/10.1103/PhysRevLett.108.051302}{\emph{Phys. Rev.
  Lett.} {\bf 108} (2012) 051302}, [\href{http://arxiv.org/abs/1110.3230}{{\tt
  1110.3230}}].

\bibitem{Bellini:2013lnn}
{\scshape Borexino} collaboration, G.~Bellini et~al., \emph{{Final results of
  Borexino Phase-I on low energy solar neutrino spectroscopy}},
  \href{http://dx.doi.org/10.1103/PhysRevD.89.112007}{\emph{Phys. Rev. D} {\bf
  89} (2014) 112007}, [\href{http://arxiv.org/abs/1308.0443}{{\tt 1308.0443}}].

\bibitem{Bellini:2014uqa}
{\scshape BOREXINO} collaboration, G.~Bellini et~al., \emph{{Neutrinos from the
  primary proton\textendash{}proton fusion process in the Sun}},
  \href{http://dx.doi.org/10.1038/nature13702}{\emph{Nature} {\bf 512} (2014)
  383--386}.

\bibitem{Agostini:2017cav}
{\scshape Borexino} collaboration, M.~Agostini et~al., \emph{{Improved
  measurement of $^8$B solar neutrinos with $1.5 kt·y$ of Borexino exposure}},
  \href{http://dx.doi.org/10.1103/PhysRevD.101.062001}{\emph{Phys. Rev. D} {\bf
  101} (2020) 062001}, [\href{http://arxiv.org/abs/1709.00756}{{\tt
  1709.00756}}].

\bibitem{Vinyoles:2016djt}
N.~Vinyoles, A.~M. Serenelli, F.~L. Villante, S.~Basu, J.~Bergstr\"om,
  M.~Gonzalez-Garcia et~al., \emph{{A new Generation of Standard Solar
  Models}},
  \href{http://dx.doi.org/10.3847/1538-4357/835/2/202}{\emph{Astrophys. J.}
  {\bf 835} (2017) 202}, [\href{http://arxiv.org/abs/1611.09867}{{\tt
  1611.09867}}].

\bibitem{Boehm:2013jpa}
C.~Boehm, M.~J. Dolan and C.~McCabe, \emph{{A Lower Bound on the Mass of Cold
  Thermal Dark Matter from Planck}},
  \href{http://dx.doi.org/10.1088/1475-7516/2013/08/041}{\emph{JCAP} {\bf 08}
  (2013) 041}, [\href{http://arxiv.org/abs/1303.6270}{{\tt 1303.6270}}].

\bibitem{Kannike:2020agf}
K.~Kannike, M.~Raidal, H.~Veermae, A.~Strumia and D.~Teresi, \emph{{Dark Matter
  and the XENON1T electron recoil excess}},
  \href{http://arxiv.org/abs/2006.10735}{{\tt 2006.10735}}.

\bibitem{Essig:2011nj}
R.~Essig, J.~Mardon and T.~Volansky, \emph{{Direct Detection of Sub-GeV Dark
  Matter}}, \href{http://dx.doi.org/10.1103/PhysRevD.85.076007}{\emph{Phys.
  Rev. D} {\bf 85} (2012) 076007}, [\href{http://arxiv.org/abs/1108.5383}{{\tt
  1108.5383}}].

\bibitem{Essig:2015cda}
R.~Essig, M.~Fernandez-Serra, J.~Mardon, A.~Soto, T.~Volansky and T.-T. Yu,
  \emph{{Direct Detection of sub-GeV Dark Matter with Semiconductor Targets}},
  \href{http://dx.doi.org/10.1007/JHEP05(2016)046}{\emph{JHEP} {\bf 05} (2016)
  046}, [\href{http://arxiv.org/abs/1509.01598}{{\tt 1509.01598}}].

\bibitem{Akerib:2019fml}
{\scshape LZ} collaboration, D.~S. Akerib et~al., \emph{{The LUX-ZEPLIN (LZ)
  Experiment}}, \href{http://dx.doi.org/10.1016/j.nima.2019.163047}{\emph{Nucl.
  Instrum. Meth. A} {\bf 953} (2020) 163047},
  [\href{http://arxiv.org/abs/1910.09124}{{\tt 1910.09124}}].

\bibitem{Ade:2015xua}
{\scshape Planck} collaboration, P.~Ade et~al., \emph{{Planck 2015 results.
  XIII. Cosmological parameters}},
  \href{http://dx.doi.org/10.1051/0004-6361/201525830}{\emph{Astron.
  Astrophys.} {\bf 594} (2016) A13},
  [\href{http://arxiv.org/abs/1502.01589}{{\tt 1502.01589}}].

\bibitem{Yuksel:2007xh}
H.~Yuksel, J.~F. Beacom and C.~R. Watson, \emph{{Strong Upper Limits on Sterile
  Neutrino Warm Dark Matter}},
  \href{http://dx.doi.org/10.1103/PhysRevLett.101.121301}{\emph{Phys. Rev.
  Lett.} {\bf 101} (2008) 121301}, [\href{http://arxiv.org/abs/0706.4084}{{\tt
  0706.4084}}].

\bibitem{Boyarsky:2007ge}
A.~Boyarsky, D.~Malyshev, A.~Neronov and O.~Ruchayskiy, \emph{{Constraining DM
  properties with SPI}},
  \href{http://dx.doi.org/10.1111/j.1365-2966.2008.13003.x}{\emph{Mon. Not.
  Roy. Astron. Soc.} {\bf 387} (2008) 1345},
  [\href{http://arxiv.org/abs/0710.4922}{{\tt 0710.4922}}].

\bibitem{Essig:2013goa}
R.~Essig, E.~Kuflik, S.~D. McDermott, T.~Volansky and K.~M. Zurek,
  \emph{{Constraining Light Dark Matter with Diffuse X-Ray and Gamma-Ray
  Observations}}, \href{http://dx.doi.org/10.1007/JHEP11(2013)193}{\emph{JHEP}
  {\bf 11} (2013) 193}, [\href{http://arxiv.org/abs/1309.4091}{{\tt
  1309.4091}}].

\bibitem{Perez:2016tcq}
K.~Perez, K.~C.~Y. Ng, J.~F. Beacom, C.~Hersh, S.~Horiuchi and R.~Krivonos,
  \emph{{Almost closing the \ensuremath{\nu}MSM sterile neutrino dark matter
  window with NuSTAR}},
  \href{http://dx.doi.org/10.1103/PhysRevD.95.123002}{\emph{Phys. Rev. D} {\bf
  95} (2017) 123002}, [\href{http://arxiv.org/abs/1609.00667}{{\tt
  1609.00667}}].

\bibitem{Roach:2019ctw}
B.~M. Roach, K.~C.~Y. Ng, K.~Perez, J.~F. Beacom, S.~Horiuchi, R.~Krivonos
  et~al., \emph{{NuSTAR Tests of Sterile-Neutrino Dark Matter: New Galactic
  Bulge Observations and Combined Impact}},
  \href{http://dx.doi.org/10.1103/PhysRevD.101.103011}{\emph{Phys. Rev. D} {\bf
  101} (2020) 103011}, [\href{http://arxiv.org/abs/1908.09037}{{\tt
  1908.09037}}].

\end{thebibliography}\endgroup

\end{document}